\documentclass[prb,twocolumn,showpacs,preprintnumbers,amsmath,amssymb]{revtex4}
\usepackage{CJK}
\usepackage{graphicx}
\usepackage{amssymb}
\usepackage[latin1]{inputenc}
\usepackage{dcolumn}
\usepackage{bm}
\usepackage{amsmath}
\usepackage{mathtools}
\usepackage{textcomp}
\usepackage{amsbsy}
\usepackage[dvips]{color}
\usepackage{float}
\usepackage{times}
\usepackage{bbm}
\DeclareMathAlphabet\mathbfcal{OMS}{cmsy}{b}{n}
\DeclareGraphicsExtensions{.png,.eps}
\usepackage{mathrsfs}
\usepackage{amssymb}
\usepackage{wasysym}
\usepackage{extarrows}
\usepackage{cleveref}
\usepackage{float}
\usepackage{subfig}
\usepackage{dsfont}
\usepackage{cases}
\usepackage{rotating}
\usepackage{placeins}
\usepackage[dvipsnames]{xcolor}

\begin{document}

\begin{CJK*}{GB}{gbsn}



\title{Quantum Geometric Exciton Drift Velocity}

\author{Jinlyu Cao$^{1,2}$, H.A.Fertig$^{1,2}$, and Luis Brey$^3$}
\affiliation{
$^1$ Department of Physics, Indiana University, Bloomington, IN 47405\\
$^2$ Quantum Science and Engineering Center, Indiana University, Bloomington, IN, 47408\\
$^3$ Instituto de Ciencia de Materiales de Madrid, (CSIC),
Cantoblanco, 28049 Madrid, Spain\\
}

\date{\today}


\begin{abstract}
In many situations, excitons -- bound particle-hole pairs above an insulating ground state -- carry an electric dipole moment, allowing them to be manipulated via coupling to an electric field.  For two-dimensional systems we demonstrate that this property of an exciton is uniquely determined by the quantum geometry of its eigenstates, and demonstrate its intimate connection with a quantity which we call the {\it quantum geometric dipole.}  We demonstrate that this quantity arises naturally in the semiclassical equations of motion of an exciton in an electric field, adding a term additional to the anomalous velocity coming from the Berry's curvature.  In a uniform electric field this contributes a drift velocity to the exciton akin to that expected for excitons in crossed electric and magnetic fields, even in the absence of a real magnetic field.  We compute the quantities relevant to semiclassical exciton dynamics for several interesting examples of bilayer systems with weak interlayer tunneling and Fermi energy in a gap, where the exciton may be sensibly described as a two-body problem.  These quantities include the exciton dispersion, its quantum geometric dipole, and its Berry's curvature. For a simple example of two gapped-graphene layers in vanishing magnetic field, we demonstrate that there is a non-vanishing quantum geometric dipole when the layers are different, e.g., have different gaps, but vanishes when the layers are identical.  We further analyze examples in the presence of magnetic fields, allowing us to examine cases involving graphene, in which a gap is opened by Landau level splitting. Heterostructures involving transition metal dichalcogenides materials are also considered.  In each case the quantum geometric dipole and Berry's curvatures play out in different ways.  In some cases the lowest energy exciton state is found to reside at finite momentum, with interesting possible consequences for Bose condensation in these systems.  Additionally we find situations in which the quantum geometric dipole increases monotonically with exciton momentum, suggesting that the quantum geometry can be exploited to produce photocurrents from initially bound excitons with electric fields, without the need to overcome an effective barrier via tunneling or thermal excitation.  We speculate on further possible effects of the semiclassical dynamics in geometries where the constituent layers are subject to the same or different electric fields.
\end{abstract}
\maketitle
\end{CJK*}

\widetext
\section{Introduction}
\label{sec:Introduction}
Structured van der Waals materials represent one of the richest new classes of condensed matters systems to have emerged in recent years \cite{Geim:2013aa}.  These systems are comprised of single-layer materials layered upon one another, often leading to striking behaviors that are completely absent in the constituent materials by themselves.  One of the most well-known among these is twisted bilayer graphene, in which evidence for a rich set of many-body states has been observed \cite{Lopes_2007,Bistritzer_2011,Lopes_2012,Suarez-Morell:2010aa,Cao:2018aa,Cao:2018bb,PT_review_2020}.
However,
twisted bilayer graphene is really an example of a much broader class of two-dimensional layered systems.  Beyond graphene, single layers of various materials such as transition metal dichalcogenides (TMDs) \cite{Mak:2010aa}, hexagonal boron nitride (hBN) \cite{Dean:2010aa}, or phosphorene \cite{Castellanos-Gomez:2014aa} are currently available.  By combining different such materials into heterostructures, one may search for combinations yielding useful and/or fundamentally new physical properties \cite{Gong_2014,Liu_2016,Novoselov_2016,Ajayan_2016,Duong_2017}.  Among the interesting possibilities brought into play by such systems are interlayer excitons: bound particle-hole pairs in which different constituents reside in different layers \cite{Hong_2014,Yu_2015,Yu_2015b,Chen_2016,Kozawa_2016,Latini_2017,Wu_2018,Kunstmann_2018,Gillen_2018,Torun_2018,Ovesen_2019,Ciarrocchi_2019,Gerber_2019,Kamban_2020,Editor:2020aa}.  Because of the physical separation of the electron and hole constituents, they  can persist in these structures for very long time scales ($>100$ nsec), allowing their evolution over time to be detected optically \cite{Rivera_2015,Rivera_2016,Gao_2017,Ross_2017,Nagler_2017,Jauregui_2019}.  In some situations the excitons may Bose condense \cite{Wang:2019aa}, or organize into periodic arrays
with potential opto-electronic applications \cite{Yu_2017,Tran_2019,Seyler_2019,Jin_2019,Alexeev:2019aa,Jin:2019aa,Guo_2020}.

A fundamentally interesting aspect of excitons is that they are perhaps the simplest many-body system where quantum geometric phases can have an impact.  This arises in two ways: because the electron and hole each reside in different bands, the exciton as a composite object can bring the quantum geometries of the two bands together in a non-trivial way.  For example, the band Berry's curvatures and the quantum geometric tensors \cite{Garate:2011aa,Srivastava_2015,Zhou_2015,Cao:2018bb,Zhang:2018aa} can have important consequences for exciton energetics.  Beyond this, an exciton state is typically labeled by a total momentum, through which one may define geometric phases for the exciton as a collective object.  These in turn impact the exciton equations of motion in external electric and magnetic fields \cite{Chang_1996,Yao_2008,Kuga_2008,Qiu:2015aa,Wu_2017,Kwan_2020}, in principle allowing one to detect their presence.

\begin{figure}[!tbp]
  \centering
{\includegraphics[width=0.35\textwidth]{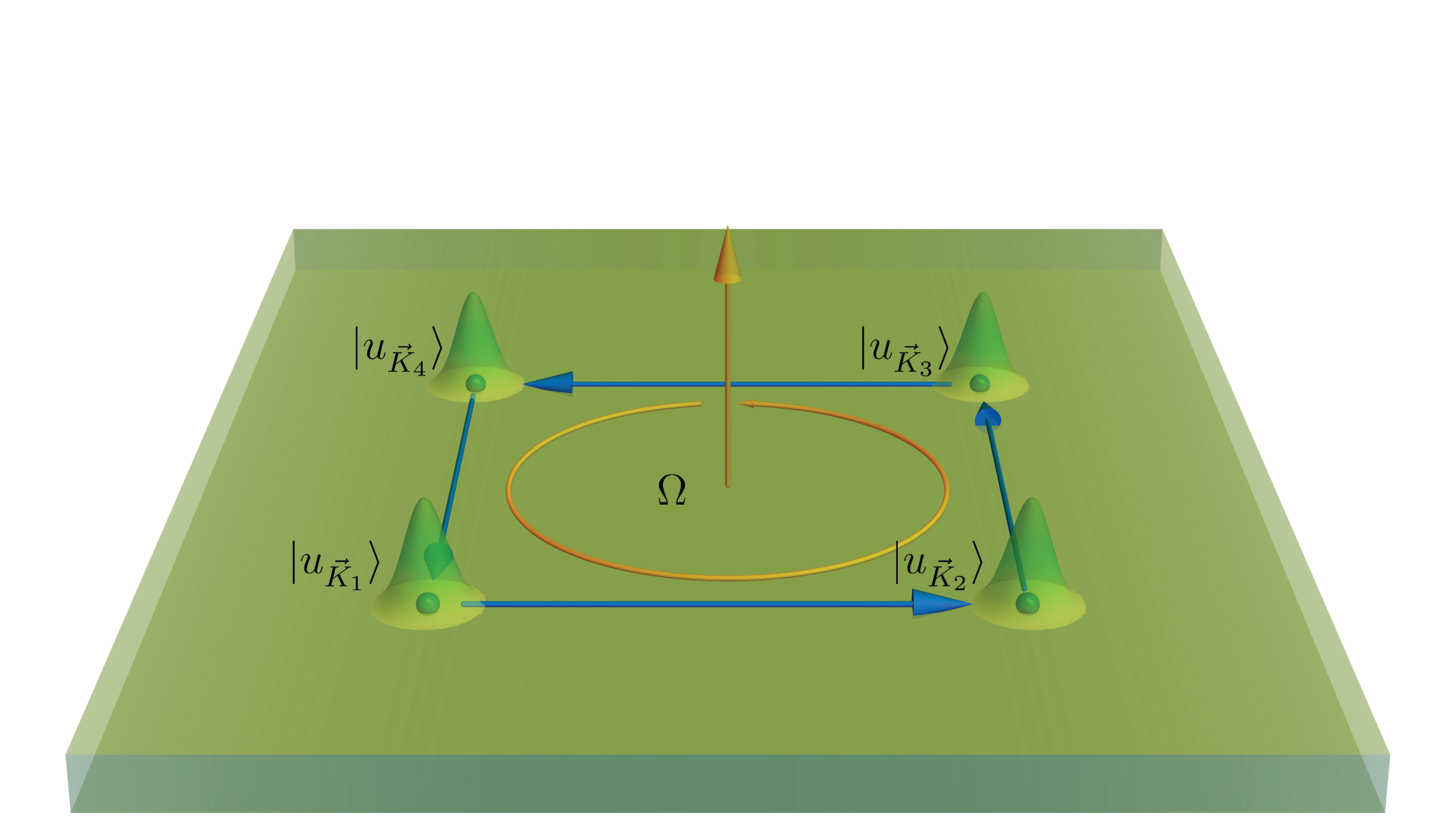}}
  \hfill
 {\includegraphics[width=0.60\textwidth]{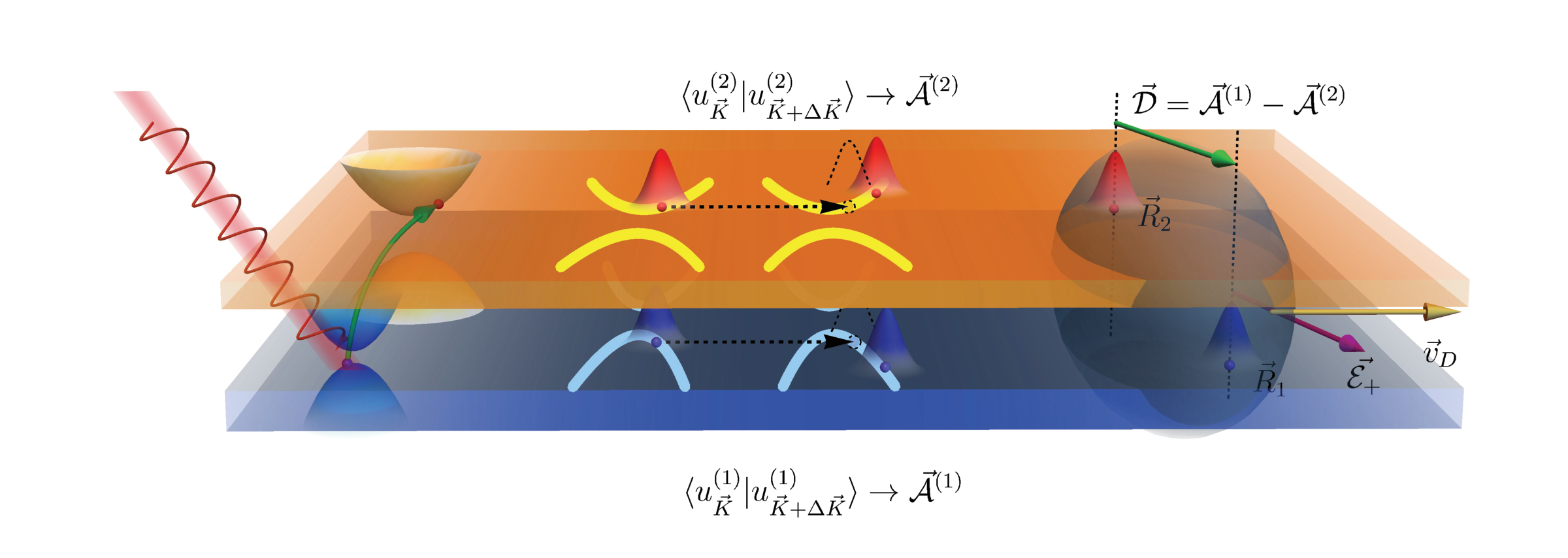}}
      \caption{(a) Berry's curvature for a single particle, such as an electron in a band structure, is usually characterized by the evolution of a state as it is carried around a loop in its parameter space, and can be understood as an effective flux through that loop.  (b) For two-particle systems (such as
      an exciton -- a bound state of a hole and electron, often generated by laser excitation, illustrated for a bilayer system on left) supports a different quantum geometric measure, based on connections ${\mathbfcal A}^{(1)}$ (${\mathbfcal A}^{(2)}$) that characterize changes in the quantum state of the system as the total momentum ${\bf K}$ changes by a small amount ($\Delta {\bf K}$), with plane wave phases assigned completely to the hole (bottom, middle) or the electron (top, middle).  The {\it difference} in these connections defines
      ${\mathbfcal D}$, the {\it quantum geometric dipole}, which in turn dictates the exciton dipole moment. For an exciton wavepacket, when a net electric field ${\mathbfcal E}_+$ has a component parallel to ${\mathbfcal D}$, a perpendicular drift velocity ${\bf v}_D$ typically develops (right), dynamics analogous to that expected in a magnetic field, even when is no such field.}
  \label{wavepacket_cartoon}
\end{figure}

In this paper, we will introduce a new quantum geometric quantity for two-dimensional systems which we call the {\it quantum geometric dipole}.  The difference between this quantity and the more commonly studied Berry's curvature for single electrons is illustrated in Fig. \ref{wavepacket_cartoon}. The quantum geometric dipole is intrinsically two-body in nature, and its physical manifestation for an exciton appears in its electric dipole moment.  In what follows we develop this concept, and discuss its realization in interlayer excitons of several bilayer heterostructure systems, focusing on systems where tunnel coupling between the materials is weak.  These turn out to be attractive environments for studying the quantum geometric dipole for a number of reasons.  Among these is that they allow one to explore the effects of differing quantum geometric environments for each of the constituents.  (Indeed we will see that in situations involving layers of the same material, the interesting contributions to the quantum geometric dipole can cancel away.) An extreme example of this is the case where one of the  materials is graphene, for which the band Berry's curvature vanishes, while the other material does have single-particle curvature.  In this case it is important to impose a magnetic field so that gaps open in the graphene spectrum, admitting a well-defined two-body approach for the exciton problem.  Beyond this, the relative isolation of the two layers allows in principle for different electric fields imposed within the layers, resulting in some control over their dynamics.  Finally, when tunneling may be neglected, many-body effects become relatively unimportant, and the excitons may be accurately treated as a two-body system.

In developing this concept, most of the examples we discuss below involve systems with a magnetic field perpendicular to the layers.  In fact,
the connection between the dipole moment of an exciton in a strong magnetic and its momentum has been known for some time \cite{Kallin_1984};  however, to our knowledge, its quantum geometric origin has not been recognized.  Because of this, as we show below, the connection between momentum and dipole moment in an exciton is much more general than the strong field problem suggests.  We will show that the quantum geometric dipole dictates a dipole moment as a function of exciton momentum in a model heterostructure of gapped graphene in {\it zero} magnetic field, as well as for magneto-excitons in graphene/TMD heterostructures, and TMD/TMD heterostructures.


The physical importance of the exciton dipole moment becomes apparent when one considers the dynamics of an exciton in an electric field $\mathbfcal{E}$, to which the dipole moment couples, which we demonstrate below by analyzing semiclassical equations of motion.  In agreement with previous studies we find that the exciton energy dispersion and Berry's curvature both enter these, but, importantly, in addition we find a contribution that couples to the quantum geometric dipole.  For a uniform electric field, a non-vanishing quantum geometric dipole/moment leads to a constant drift velocity, which in the strong magnetic field case takes the form ${\bf v}_D = \mathbfcal{E} \times {\mathbf B} /B^2$, where in the frame of reference of the exciton the electric field vanishes \cite{Lukose_2007,Jackson_book}.
Beyond the strong field limit, however, the band environments modify the dipole moment, so that the drift in a uniform electric field behaves as if the magnetic field in which it is moving is {\it different} than what is externally applied.  Indeed, we will see that even in the absence of an external applied field, a quantum geometric dipole leads to drift motion.  The
basic effect is illustrated in Fig. \ref{wavepacket_cartoon} for an exciton wavepacket in a uniform in-plane electric field ${\mathbfcal E}$.

The presence of geometric phases may thus cause behavior rather similar to that expected of a system in a magnetic field, as has been pointed out for single particles projected into individual bands with non-vanishing Chern numbers \cite{Qi_2011}.  In principle, these effects may be detected by creating excitons at a fixed location of a heterostructure, and using photoluminescence to look for drift of excitons induced by the electric field away from the excitation location.  The interlayer excitons could be induced via a localized laser spot, for which in-plane particle-hole pairs may relax into lower energy inter-layer excitons through a small residual tunneling.  We also propose using an in-plane laser beam with electric field linearly-polarized out-of-plane, to minimize photoluminescence from in-plane excitons which may mask the signals from the excitons of interest.

\begin{figure}
  \centering
  \includegraphics[width=0.9\textwidth,trim= 0 0 0 0,clip]{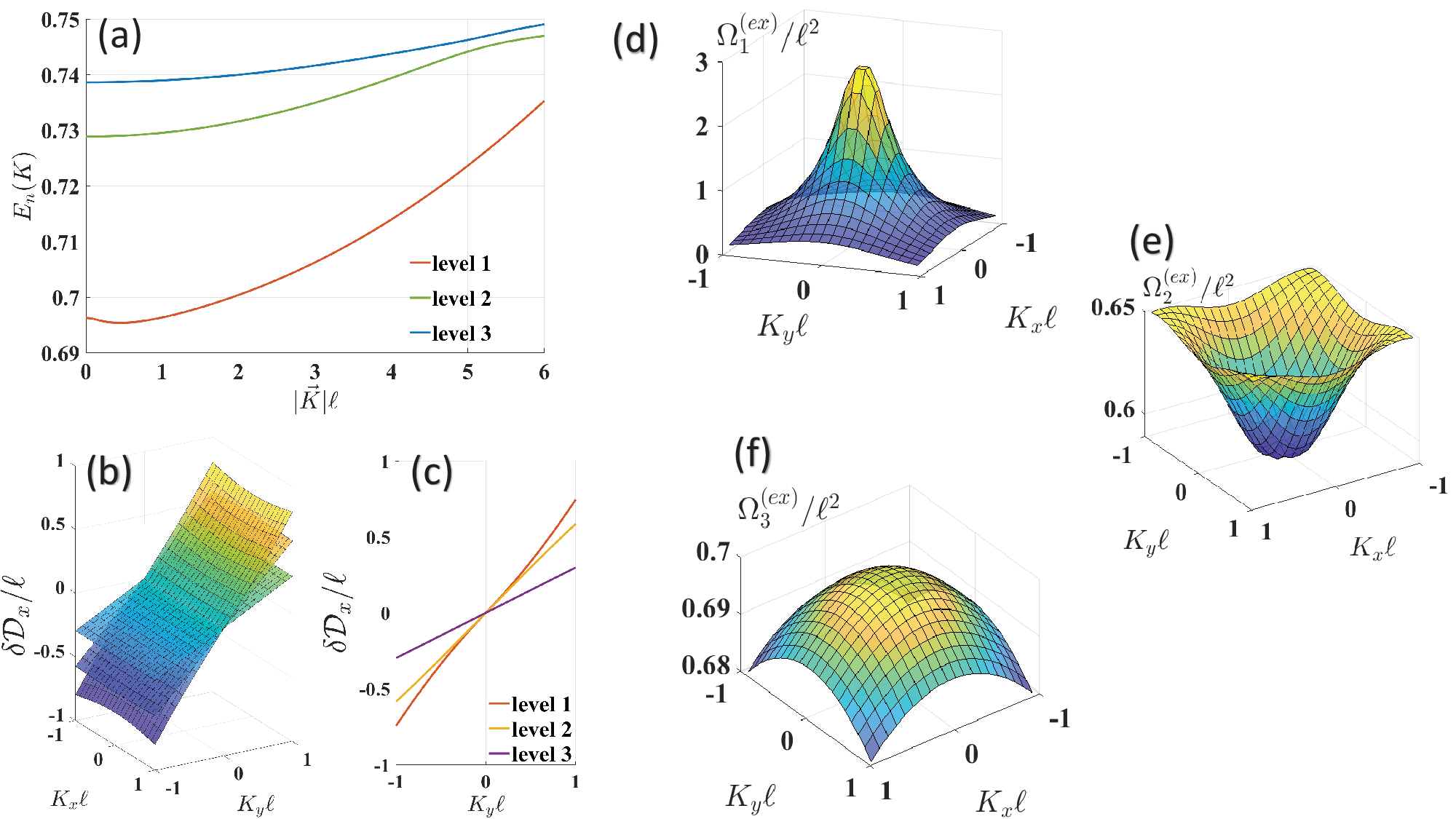}
  \caption{Properties of magneto-exciton in a graphene/TMD heterostructure. Hole is assumed to be in the $K'$ valley of the TMD. (a) Energy dispersions for three lowest levels (in units of eV) as a function of exciton momentum $|{\bf K}|$. (b) $\hat{x}$-component of the addition to the quantum geometric dipole beyond the single Landau level result, $\delta {\mathcal D}_x$, as a function of ${\bf K}$.  The full quantum geometric dipole is given by ${\mathbfcal D}= \delta {\mathbfcal D} + {\bf K} \times \hat{z} \ell^2$. (c) $\delta {\mathcal D}_x$ vs. exciton momentum ${\bf K}$ for $K_{x}=0$.  (d), (e), (f):
  Berry's curvatures of low-lying excitons as a function of momentum ${\bf K}$ for three lowest exciton levels. TMD parameters chosen as appropriate for MoS$_2$ \cite{Xiao_2012}.  Magnetic field is taken as $B=10$ T, layer separation is 0.71 nm.  A dielectric constant of 4 is assumed for the Coulomb interaction.    Calculations performed retaining Landau levels with index $|n| \le 50$ for the hole and $|n| \le 50$ for the electron.  }
  \label{G_MoS2_K_results}
\end{figure}
Figure \ref{G_MoS2_K_results} illustrates representative results of our analysis of quantities entering the equation of motion, for magneto-excitons in a simple model of a graphene/TMD heterostructure.  The change in the exciton dipole moment when the full quantum geometric structure is accounted for -- beyond the single Landau level approximation -- is considerable.  These quantities also vary among different exciton modes, suggesting that they might be spatially segregated by  electric fields.  Beyond this, we find, within our long-wavelength description, that the dipole moments of magneto-excitons grow without bound as the exciton momentum increases, so that electron and hole may be continuously separated by an electric field that is {\it anti}-symmetric in layer.  Such an effect could be detected as a photocurrent flowing perpendicular to the applied field.  It is also notable that in this system, the lowest energy exciton state appears at finite magnitude of the exciton momentum $K$, suggesting that a large density of these at low temperature could form a condensate with a broken rotational spatial symmetry, forming an additional ordering beyond that of the superfluid order parameter expected of an exciton condensate.

This article is organized as follows.  In Section \ref{section:LandauLevels}, we demonstrate the basic idea of a quantum geometric dipole for an exciton composed of an electron and hole in different Landau levels.  The analysis reproduces the known result \cite{Kallin_1984} for the exciton dipole moment, while demonstrating its quantum geometric nature.  We describe in Section \ref{section:Dynamics} the types of model systems considered in our study, and then derive semiclassical equations for excitons, demonstrating the direct way in which the quantum geometric dipole enters.  This is followed in Section \ref{section:gapped_graphene} by an analysis of the quantum geometric dipole and Berry's curvature for a simple model of two gapped graphene layers in zero magnetic field, demonstrating that a real field is not needed to produce the dipole physics.  Section \ref{section:Heterostructures} presents our numerical results for magneto-excitons in graphene/TMD and TMD/TMD heterostructures.  We conclude in Section \ref{section:summary_and_discussion} with a summary of our results, as well as some discussion of possible geometries in which the effects discussed in this work might be observed, and of further interesting physics suggested by our studies.  This article also has three appendices.  Appendix \ref{appendix:lagrangian} describes some details of our derivation of the semiclassical equations of motion; Appendix \ref{appendix:exciton_equations} provides some details of how we compute exciton wavefunctions and energies, as well as Berry's connections, quantum geometric dipoles, and Berry's curvatures; and Appendix \ref{appendix:exciton_generation} discusses exciton generation by light with electric fields linearly polarized normal to a heterostructure.














\section{Dipole Moment as a Geometric Quantity: The Case of Landau Levels}
\label{section:LandauLevels}

We begin our discussion by illustrating how the dipole moment of an exciton may be understood in terms of quantum geometric phases using a relatively simple example, where the exciton dipole moment is already known: magnetoexcitons in a strong magnetic field, where the quantum states are well-described by an electron and a hole each residing in a single Landau level.  For a single particle of charge $se$ ($s=\pm 1$), the Hamiltonian is given by
\begin{equation}
\label{H1body}
H_{1-body}^{(s)}({\bf r}) = {1 \over {2m^*}} \left({\mathbf p}-se {\mathbf A}\right)^2
\end{equation}
where $m^*$ is the effective mass of the carrier, $e > 0$ is the electron charge, $\mathbf{p}=-i\vec\nabla$ (with $\hbar=1$), and the vector potential $\mathbf{A}$ is taken here to be $B(-y/2,x/2,0)$, with $B$ a perpendicular magnetic field.
Eigenstates of this Hamiltonian take the form
\begin{equation}
\label{LLstates}
\langle {\bf r} |\psi_{n,k}^{(s)}\rangle \equiv \psi_{n,k}^{(s)} (x,y) = \frac{1}{\sqrt{\ell \sqrt{\pi} 2^n n! L_y}}
e^{-isxy/2\ell^2}
e^{i k y} e^{-\frac{(x- s k \ell^2)^2}{2 \ell^2} }H_n(\frac{x - s k \ell^2}{\ell}),
\end{equation}
where the non-negative integer $n$ is the Landau level index, $H_n$ is a Hermite polynomial, and $\ell=\sqrt{\hbar/eB}$ is the magnetic length. The energies of these states are $\varepsilon_n=(n+{1 \over 2})\omega_c$, where $\omega_c=eB/m^*$ is the cyclotron frequency.  Because these energies are independent of $k$, each Landau level has a large degeneracy, $L_xL_y/2\pi \ell^2$, where $L_xL_y$ is the area of the sample.  We assume the electron density is such that, in the ground state, every Landau level is either completely filled or completely empty, so that, in the strong field limit, low-energy excitations involve removing an electron from a filled Landau level and placing it in an empty one.  In this situation one may reasonably model the excitation as a two-body system, roughly analogous to a positronium ``atom'' in a strong magnetic field \cite{Bychkov_1981,Kallin_1984}.

To identify the momentum of the excitation it is convenient to recast the two-body states in terms of eigenstates of magnetic translations \cite{Zak_1964}.  To do this, we define real-space translation operators along the $\hat{x}$ ($\hat{y}$) direction of displacement $a_x$ ($a_y$).  For a charge of sign $s$ they take the form \cite{Zak_1964}
\begin{align}\label{e_MTO}
  \hat{T}^{(s)}_{a_x\hat{x}} = e^{i (p_x - s\frac{y}{2\ell^2})a_x},\\
  \hat{T}^{(s)}_{a_y\hat{y}} = e^{i (p_y + s\frac{x}{2\ell^2})a_y}.
\end{align}
Writing the magnetic flux through a unit cell as $\Phi_{uc}$, and the magnetic flux quantum as $\Phi_0 \equiv e^2/h$, provided one chooses $\Phi_{uc}/\Phi_0 \equiv a_xa_y/2\pi\ell^2$ to be an integer, these operators commute with one another, as well as with the single particle Hamiltonians for the electron and hole (Eq. \ref{H1body}).  One may then construct eigenstates of all three operators.  For $\Phi_{uc}/\Phi_0=1$ these are
\begin{equation}
\label{ewfsc}
\langle {\bf r}|n,{\bf k},s\rangle \equiv \phi^{(s)}_{n,{\bf k}}(x,y)=\left(\frac{a_y}{L_y}\right)^{1/2}\sum_{m} e^{isk_x a_x m} \psi_{n,k_y+m\Delta k}^{(s)},
\end{equation}
with $\Delta k=2\pi/a_y$.  The eigenvalues of these states under the magnetic translation $\hat{T}^{(s)}_{a_x\hat{x}}$ ($\hat{T}^{(s)}_{a_y\hat{y}}$) above are $e^{ik_xa_x}$ ($e^{ik_ya_y}$).

\subsection{Exciton States in a Strong Magnetic Field}

With the single particle states formulated in this way, states of a particle-hole pair now becomes straightforward to write down.  The Hamiltonian of the two body system is
\begin{equation}
\label{H2body}
H_{2-body}=H^{(+)}_{1-body}({\bf r}_1) + H^{(-)}_{1-body}({\bf r}_2) +v({\bf r}_1-{\bf r_2}),
\end{equation}
where ${\bf r}_1$ and ${\bf r}_2$ are the two-dimensional hole and electron positions (each in a different layer), and $v$ is an interparticle interaction.  For strong fields, the single particle energies of the Landau levels are highly separated, and we expect each particle to reside, to an excellent approximation, in a single Landau level.  If we specify the indices of these to be $n_e$ and $n_h$ for the electron and hole, respectively, we are led to consider states of the form
\begin{equation}\label{exciton_wf0}
  \langle {\bf r}_1,{\bf r}_2|\Phi_{{\bf K}}\rangle = \sum_{{\bf q}} C_{{\bf q}}({\bf K}) e^{iK_xq_y \ell^2} \phi^{(+)}_{n_h,{\bf q}}({\bf r}_1) \phi^{(-)}_{n_e,{\bf K}-{\bf q}}({\bf r}_2),
\end{equation}
where ${\bf K}$ is the exciton momentum, the sum over ${\bf q}$ is within the first Brillouin zone, specified by $-\pi/a_x \le q_x \le \pi/a_x$ and $-\pi/a_y \le q_y \le \pi/a_y$, and the phase factor $e^{iK_xq_y \ell^2}$ is introduced so that the expansion coefficients $C_{{\bf q}}({\bf K})$, and the summand as a whole, are periodic in the reciprocal lattice defined by the magnetic translations above.  The exciton states and energies are found by computing $E_{n_e,n_h}({\bf K}) \equiv \langle\Phi_{{\bf K}}|H_{2-body}|\Phi_{{\bf K}}\rangle$, and minimizing this with respect to the parameters $C_{{\bf q}}({\bf K})$.  The final result takes the form
\begin{equation}\label{exciton_wf1}
\langle {\bf r}_1,{\bf r}_2|\Phi_{{\bf K}={\bf K}^{(0)}-\hat{z} \times {\bf R}/\ell^2}\rangle = \sqrt{\frac{a_xa_y}{L_xL_y}}\sum_{{\bf q}} e^{iK_{x}^{(0)}q_y \ell^2} e^{i{\bf q}\cdot {\bf R}} \phi^{(+)}_{n_h,{\bf q}}({\bf r}_1) \phi^{(-)}_{n_e,{\bf K}^{(0)}-{\bf q}}({\bf r}_2),
\end{equation}
where  ${\bf K}^{(0)}$ lies in the first Brillouin zone of the lattice, and the sum over ${\bf q}$ is within a single Brillouin zone.
The quantity ${\bf R} = n_xa_x\hat{x} + n_y a_y\hat{y}$, with $n_x,n_y$ integers, is a point on the effective real space lattice; however, because of our choice of flux through a unit cell, $\hat{z} \times {\bf R}/\ell^2$ is a point on the reciprocal lattice, so ${\bf K}$ spans all possible values of momentum.
The corresponding energy of the exciton is
\begin{equation}
\label{single_LL_exciton_energy}
E_{n_e,n_h}({\bf K}={\bf K}^{(0)}-\hat{z} \times {\bf R}/\ell^2)=\varepsilon_{n_e}+\varepsilon_{n_h}+\int \frac{d^2 k}{(2\pi)^2} \tilde{v}({\bf k}) S^{n_h,n_h}_{0,{\bf k}}({\bf k},+)S^{n_e,n_e}_{{\bf K}^{(0)},{\bf K}^{(0)}-{\bf k}}(-{\bf k},-) e^{iK_{x}^{(0)}k_y\ell^2}e^{i{\bf k}\cdot {\bf R}}
\end{equation}
where $\tilde{v}({\bf k})$ is the Fourier transform of $v({\bf r})$.  The structure factors
\begin{equation}\label{Bloch_overlap_general}
  S^{n,n'}_{\bf{q},{q}\,'}({\bf k},s) \equiv  \langle n,{\bf q},s|e^{-i{\bf k}\cdot{\bf r}} | n',{\bf q}\,',s\rangle,
\end{equation}
play an important role here and in what follows, in that they inform the exciton of the quantum geometry of the bands -- in this case, the Landau levels -- in which the electron and hole reside \cite{Srivastava_2015,Zhou_2015,Trushin_2016,Hichri_2019}.  For the simple case of an electron and hole each in their $n=0$ Landau levels, and a Coulomb interaction between the particle and hole, $v({\bf r})=e^2/r$ (which ignores the separation between the electron and hole layers), Eq. \ref{single_LL_exciton_energy} yields the known \cite{Bychkov_1981,Kallin_1984} result $E_{0,0}(K)=\omega_c-e^2\sqrt{\frac{\pi}{2}}e^{-K^2\ell^2/4}I_0\left(\frac{K^2\ell^2}{4}\right)$, where $I_0$ is a modified Bessel function.

\subsection{Exciton Dipole Moment as a Quantum Geometric Phase}

Our interest at this moment is not so much in the energetics of an exciton as in how its state evolves when the momentum ${\bf K}$ changes slowly.  Eq. \ref{exciton_wf1} allows us to do this.  A quantitative measure of this evolution can be accomplished by looking at scalar products of states at different momenta, say $|\Phi_{{\bf K}_1}\rangle$ and $|\Phi_{{\bf K}_2}\rangle$.  However, since these are eigenstates of translation operators $\hat{T}^{(ex)}_{{\bf a}_{\mu}} \equiv \hat{T}^{(+)}_{{\bf a_{\mu}}}\hat{T}^{(-)}_{{\bf a_{\mu}}}$, with ${\bf a}_{\mu=x}$ (${\bf a}_{\mu=y}$) representing $a_x\hat{x}$ ($a_y\hat{y}$), with different eigenvalues, such scalar products will vanish.  In order to compare such wavefunctions through a scalar product one needs to strip off the phase piece of the wavefunction, to yield states that are periodic under $\hat{T}^{(ex)}_{{\bf a}_{\mu}}$, irrespective of the exciton momentum ${\bf K}$ \cite{Vanderbilt_book}.  In studies of exciton Berry's curvature, this is usually accomplished by identifying a center of mass coordinate ${\bf r}_{cm}$ (in the present case, ${\bf r}_{cm}=({\bf r}_1+{\bf r}_2)/2$, with ${\bf r}_1$ the hole position and ${\bf r}_2$ the electron position) and working with states of the form $e^{-i{\bf K} \cdot {\bf r}_{cm}} |\Phi_{\bf K}\rangle$,  usually to compute Berry's curvatures \cite{Yao_2008,Kuga_2008,Qiu:2015aa,Wu_2017,Trushin_2018,Kwan_2020}.  However, this is not the most general possibility.  As we now show, the freedom in how the unwanted phase is removed can be exploited to gain further information and insight into this type of excitation.

Towards this end, we consider states periodic under translations of the more general form
\begin{equation}
\label{udef}
|u_{\bf K},\alpha\rangle \equiv e^{-i\left[\alpha {\bf r}_1 +(1-\alpha){\bf r}_2 \right]\cdot {\bf K}} |\Phi_{\bf K}\rangle.
\end{equation}
This allows us to define Berry's connections specific to the hole and electron constituents of the exciton,
\begin{equation}
\label{alpha_berry}
{\mathbfcal A}^{(1)}({\bf K}) \equiv i\langle u_{\bf K},\alpha=1|\vec{\nabla}_K|u_{\bf K},\alpha=1\rangle, \quad {\mathbfcal A}^{(2)}({\bf K}) \equiv i\langle u_{\bf K},\alpha=0|\vec{\nabla}_K|u_{\bf K},\alpha=0\rangle.
\end{equation}
These quantities can be directly related to the dipole moment of an exciton state, ${\bf d} \equiv e \langle \Phi_{\bf K} | {\bf r}_1 - {\bf r}_2|\Phi_{\bf K}\rangle$.  We recast this in the form
\begin{align}
\label{dipole1}
\bf{d} &=  i e \left[\langle
\Phi_{\bf K} | e^{i{\bf K}\cdot{\bf r_1}} \left( \vec{\nabla}_K e^{-i{\bf K}\cdot{\bf r_1}} \right) |\Phi_{\bf K}\rangle
- \langle \Phi_{\bf K} | e^{i{\bf K}\cdot{\bf r_2}}
\left( \vec{\nabla}_K e^{-i{\bf K}\cdot{\bf r_2}} \right) | \Phi_{\bf K} \rangle
\right] \\
\label{dipole1b}
&=  i e \left[\langle u_{\bf K},1 |\vec{\nabla}_K | u_{\bf K},1 \rangle - \langle u_{\bf K},0 |\vec{\nabla}_K | u_{\bf K},0 \rangle \right] \\
\label{dipole1c}
&\equiv e\left[{\mathbfcal A}^{(1)}({\bf K})- {\mathbfcal A}^{(2)}({\bf K}) \right].
\end{align}
In this formulation one can see that the dipole moment is related in a direct way to a quantum geometric property of excitons with well-defined momentum.  To stress this point, from hereon we call the quantity ${\mathbfcal D}({\bf K}) \equiv {\mathbfcal A}^{(1)}({\bf K})- {\mathbfcal A}^{(2)}({\bf K})$ the {\it quantum geometric dipole.}  This result is very general: it applies to any collection of exciton states that can be labeled by momentum. While the quantities entering its definition, ${\mathbfcal A}^{(1)}({\bf K})$, ${\mathbfcal A}^{(2)}({\bf K})$, are gauge-dependent (i.e., under $|\Phi_{\bf K} \rangle \rightarrow e^{i\phi({\bf K})} |\Phi_{\bf K} \rangle$ their values change), their difference -- and so the quantum geometric dipole -- is easily seen to be gauge-invariant. This must be the case since $\bf{d}$ is in principle measurable.

To illustrate, we compute this quantity for our example of an electron-hole pair in a strong magnetic field.  Because  we have an analytic form of the wavefunction (Eq. \ref{exciton_wf1}), the computation can be carried out explicitly.  A key quantity in doing so is the overlap
\begin{equation}
\label{overlap}
\Gamma_{\alpha}({\bf K}_1,{\bf K}_2) \equiv \langle u_{{\bf K}_1},\alpha|u_{{\bf K}_2},\alpha\rangle,
\end{equation}
in terms of which we can write
\begin{equation}
\label{Gamma_to_A}
{\mathbfcal A}^{(1)}({\bf K}_1)- {\mathbfcal A}^{(2)}({\bf K}_1) = i\lim_{{\bf K}_2 \rightarrow {\bf K}_1} \vec{\nabla}_{K_2} \left[\Gamma_1({\bf K}_1,{\bf K}_2)-\Gamma_0({\bf K}_1,{\bf K}_2) \right].
\end{equation}
Since $\Gamma_1$ ($\Gamma_0$) involves a plane wave factor $e^{i\left({\bf K}_1- {\bf K}_2\right) \cdot {\bf r}_1}$ ($e^{i\left({\bf K}_1- {\bf K}_2\right) \cdot {\bf r}_2}$) for the hole (electron) only, these can be written directly in terms of the structure factors in Eq. \ref{Bloch_overlap_general},
\begin{equation}
\label{Gamma1}
\Gamma_{1}({\bf K}_1,{\bf K}_2) = \frac{a_xa_y}{L_xL_y} \sum_{\bf q}
e^{-iK_{1x}^{(0)}(q_y-K_{2y}^{(0)})\ell^2+iK_{2x}^{(0)}(q_y-K_{1y}^{(0)})\ell^2+i({\bf K}_2^{(0)}-{\bf K}_1^{(0)}) \cdot {\bf R}}
S^{n_h,n_h}_{{\bf q}-{\bf K}_2^{(0)},{\bf q}-{\bf K}_1^{(0)}}({\bf K}_2^{(0)}-{\bf K}_1^{(0)},+)
\end{equation}
and
\begin{equation}
\label{Gamma0}
\Gamma_{0}({\bf K}_1,{\bf K}_2) = \frac{a_xa_y}{L_xL_y} \sum_{\bf q}
e^{ i(K_{2x}^{(0)} - K_{1x}^{(0)})q_y\ell^2}
S^{n_e,n_e}_{{\bf K}_1^{(0)}-{\bf q},{\bf K}_2^{(0)}-{\bf q}}({\bf K}_2^{(0)}-{\bf K}_1^{(0)},-).
\end{equation}
In writing these we have taken ${\bf K}_{1,2}={\bf K}_{1,2}^{(0)}-\hat{z} \times {\bf R}/\ell^2$, where ${\bf K}_{1,2}^{(0)}$ lies in the first exciton Brillouin zone, and we have assumed any additional reciprocal lattice vectors $\hat{z} \times {\bf R}/\ell^2$ needed to specify the momenta are the same for both, since ultimately we will set ${\bf K}_2 \rightarrow {\bf K}_1$.  The structure factors (Eq. \ref{Bloch_overlap_general}) are somewhat tedious to work out, but their forms may eventually be written compactly as
\begin{align}
\label{structure_factor_hole}
S^{n_h,n_h}_{{\bf q}-{\bf K}_2^{(0)},{\bf q}-{\bf K}_1^{(0)}}(\Delta {\bf K},+) &=
e^{i\Delta K_x \left({1 \over 2} K_{2y}^{(0)} + {1 \over 2}K_{1y}^{(0)} -q_y \right)\ell^2-\Delta K^2\ell^2/4} L_{n_h}\left( \frac{\Delta K^2\ell^2}{2} \right) ,
\\
\label{structure_factor_electron}
S^{n_e,n_e}_{{\bf K}_1^{(0)}-{\bf q},{\bf K}_2^{(0)}-{\bf q}}(\Delta {\bf K},-) &= e^{i\Delta K_x \left({1 \over 2} K_{1y}^{(0)} + {1 \over 2} K_{2y}^{(0)}-q_y \right) \ell^2- \Delta K^2\ell^2/4} L_{n_e}\left( \frac{\Delta K^2\ell^2}{2} \right),
\end{align}
where $\Delta {\bf K} \equiv {\bf K}_2-{\bf K}_1$, and $L_n$ are Laguerre polynomials.
Substituting Eqs. \ref{structure_factor_hole} and \ref{structure_factor_electron} into Eqs. \ref{Gamma1} and \ref{Gamma0}, we arrive at
\begin{equation}
\label{Gamma_diff}
\Gamma_{1}({\bf K}_1,{\bf K}_2) - \Gamma_{0}({\bf K}_1,{\bf K}_2) =
e^{ -\Delta K^2\ell^2/4}\left[e^{{i \over 2} \left(K_{1x}^{(0)}+K_{2x}^{(0)}\right)\Delta K_y\ell^2 + i\Delta {\bf K} \cdot {\bf R}} L_{n_h}\left( \frac{\Delta K^2\ell^2}{2} \right)
- e^{{i \over 2} \left(K_{1y}^{(0)}+K_{2y}^{(0)}\right)\Delta K_x\ell^2 }L_{n_e}\left( \frac{\Delta K^2\ell^2}{2} \right) \right].
\end{equation}


Finally, using Eq. \ref{Gamma_to_A}, and remembering ${\bf K} = {\bf K}^{(0)}-\hat{z} \times {\bf R}/\ell^2$ one finds the quantum geometric dipole to be
\begin{equation}
\label{dipole_curv_LL}
{\mathbfcal D} ({\bf K})= {\mathbfcal A}^{(1)}({\bf K})- {\mathbfcal A}^{(2)}({\bf K}) = {\bf K} \times \hat{z} \ell^2,
\end{equation}
so that the electric dipole moment of the exciton is just ${\bf d}= e {\bf K} \times \hat{z} \ell^2.$  Note the result is independent of the specific Landau levels $n_h$ and $n_e$ in which the hole and the electron reside.

The connection between the dipole moment of a magnetoexciton in the strong field limit and its momentum is well-known \cite{Kallin_1984}; the present analysis however demonstrates its quantum geometric nature.  The presence of this dipole moment may be understood semi-classically as a balancing of the electric attraction between the electron and the hole, and the magnetic force that results from the uniform motion of the exciton in the magnetic field.  We will see more generally that excitons carry dipole moments dictated by the quantum geometric nature of their wavefunctions as a function of their momentum ${\bf K}$.  Before showing this explicitly, we consider how their presence impacts that {\it dynamics} of excitons.

\section{Model Systems and Semiclassical Exciton Dynamics}
\label{section:Dynamics}
In this section we derive equations of motion for an exciton wavepacket, in a way that is relevant for interlayer excitons in a two-layer system.  In particular we consider the possibility that there may be electric fields in the individual layers that are different, a situation that becomes possible when interlayer tunneling is only a small perturbation, which is the limit we have focused upon throughout this paper.  The method we follow was introduced in Ref. \onlinecite{Chang_1996} and has previously been applied to excitons \cite{Yao_2008}.  We will see that by focusing on the electron and hole coordinates separately one arrives at a somewhat different result, in which the quantum geometric dipole enters in a direct and physically intuitive way.  Before beginning this discussion, however, we need to specify more concretely the forms of the single-particle Hamiltonians of the exciton constituents, for the types of systems which are the main focus of our study.

\subsection{Single Layer Hamiltonians}
\label{noninteracting_models}
The models we discuss below involve single, two-dimensional layers in which an electron or hole resides, which may be graphene, gapped graphene, or a TMD system.
Without a magnetic field, at long wavelengths the wavefunctions for each of these can be described by two-component spinors.  In graphene these represent amplitudes for electrons to reside in $p_z$ orbitals on the $A$ and $B$ sublattices of the honeycomb lattice; in TMD materials the amplitudes are for different combinations of $d$ orbitals residing on the transition metal sites \cite{Xiao_2012}.
In all these cases the Hamiltonians can be cast into the general form
\begin{equation}\label{eq: single layer Hamiltonian general form}
  H^{Dirac}_e({\bf p}) = \vec{V}_{\tau} \cdot {\bf p} + m_{\sigma\tau} \sigma_{z},
\end{equation}
where each component of $\vec{V}_{\tau}$ is a $2 \times 2$ matrix,
$\tau= \pm 1$ is a valley index, $\sigma=\pm 1$ is a spin index, $m_{\sigma\tau}$ is an effective
mass parameter,
and $\sigma_z$ is a Pauli matrix that acts on the two-component spinor.
The Hamiltonian of electrons in gapped graphene reads
\begin{equation} \label{G_Hamiltonian_0}
  H _{G}^{\tau}({\bf p}) = v_G \tau \sigma_{x} p_x+ v_G \sigma_{y} p_y + \delta \sigma_z,
\end{equation}
where $v_G$ is the Fermi velocity for electrons and the mass parameter $\delta$ opens a gap in the spectrum.  Physically this can occur if the graphene resides on a substrate that results in different potentials on the $A$ and $B$ sublattices, which is effectively the case, for example, for graphene on hBN with the symmetry axes of the two materials approximately aligned \cite{Giovannetti_2007,Hunt_2013}.  When misaligned, or on a substrate a with significantly different lattice parameter, the gap parameter $\delta$ vanishes, yielding the well-known gapless Dirac point spectrum \cite{Castro_Neto_2009}.

The effective Hamiltonian for electrons in a TMD can be written in the matrix form \cite{Xiao_2012},
\begin{equation}\label{MoS2 Hamiltonian 0}
  H_{TMD}^{(\sigma,\tau)} (\vec{p}) =
  \left(
    \begin{array}{cc}
      \Delta/2 & v_F(\tau p_x - ip_y) \\
      v_F(\tau p_x + ip_y) & -\Delta/2 + \sigma\tau\lambda \\
    \end{array}
  \right),
\end{equation}
which is of the form in Eq. \ref{eq: single layer Hamiltonian general form}, up to an unimportant overall constant shift.
In this case we see the mass parameter,
$
m_{\sigma\tau} \equiv (\Delta-\sigma\tau \lambda)/2,
$
has non-trivial dependence on the spin and valley indices \cite{Xiao_2012}.

These single-particle Hamiltonians all describe electron states in the vicinity of a $K$ or $K'$ point in a band structure, and include states near a local minimum of a conduction band.  They also describe states near the top of a valence band, upon which we perform a particle-hole transformation, to describe a hole as a single positively charged particle.  The effective single-particle Hamiltonian for such a hole is given by $H^{Dirac}_h({\bf p}) = -H^{Dirac}_e({\bf -p})$.  Finally, to include the orbital effect of an external magnetic field, we make the Peierls substitution $\vec{p} \rightarrow \vec{\Pi} = \vec{p}-q\vec{A}$, where $q=-e$ for an electron and $q=+e$ for a hole.

\subsection{Semiclassical Equations of Motion}
\label{semiclassical}

The analysis for semiclassical dynamics of an exciton  begins with a wavepacket constructed from states of definite total momentum,
\begin{equation}
\label{wavepacket}
|\Psi_0\rangle \equiv \int d^2 K w({\bf K}) |\Phi_{\bf K}\rangle,
\end{equation}
where $|\Phi_{\bf K}\rangle$ is an exciton state with momentum ${\bf K}$, and $w({\bf K})$ is the normalized wavepacket weight, which is assumed to be peaked around some central momentum ${\bf K}_c$ within the first Brillouin zone, with a width that is small compared to the Brillouin zone size but not so narrow that any applied potentials vary rapidly within the real-space extent of the wavepacket \cite{Girvin_2019}.  The integration area is confined to the first Brillouin zone.  With $\langle {\bf r}_1,{\bf r}_2|\Phi_{\bf K}\rangle$ denoting the probability amplitude to find a hole at position ${\bf r}_1$ and an electron at ${\bf r}_2$, we denote the average positions for them for the wavepacket as
\begin{align}
{\bf R}_1 &\equiv \langle \Psi_0 | {\bf r}_1 | \Psi_0 \rangle, \\
{\bf R}_2 &\equiv \langle \Psi_0 | {\bf r}_2 | \Psi_0 \rangle,
\end{align}
which, using Eqs. \ref{udef} and \ref{alpha_berry}, can be recast compactly in the form
\begin{equation}
\label{position_connection}
{\bf R}_{\alpha} = \int d^2K \left[ i w^*({\bf K}) \vec{\nabla} w({\bf K})  + |w({\bf K})|^2 {\mathbfcal A^{(\alpha)}} \right]
\end{equation}
with $\alpha=1,2$.  To derive equations of motion for these quantities, we define an effective Lagrangian,
\begin{equation}
\label{lagrangian}
{\mathcal L} = \langle \Psi |i\partial_t | \Psi\rangle - \langle \Psi | H | \Psi\rangle \equiv {\mathcal L}_t - E({\bf K}),
\end{equation}
where, based on the discussion of the previous subsection,
the Hamiltonian of the system takes the form
\begin{equation}
\label{H_with_fields}
H= \sum_{i=1,2} \left\{\vec{V}^{(i)} \cdot \left[ -i\vec{\nabla}_{r_i} + (-1)^i e \left({\bf A}_0({\bf r}_i) + \delta {\bf A}^{(i)}({\bf r}_i,t) \right) \right] + M_i \right\}+ v({\bf r}_1-{\bf r}_2).
\end{equation}
Here ${\bf A}_0$ is the vector potential from a static, spatially uniform magnetic field, $-\partial_t \delta {\bf A}^{(i)} = {\mathbfcal E}^{(i)}$  specifies the electric field for the hole ($i=1$) and the electron ($i=2$), and
$\vec{\nabla} \times \delta {\bf A}^{(i)} = \delta {\mathbfcal B}^{(i)}$ encodes an additional weak magnetic field beyond that from ${\bf A}_0$.  Note that the general form of the single particle Hamiltonians discussed in the last subsection dictate how the added electric and magnetic fields are incorporated in the system.

With ${\mathcal L}$ written
terms of collective degrees of freedom, we can extremize the action $S=\int {\mathcal L}dt$ in terms of them to obtain equations of motion.  (We present some details of the how the Lagrangian is found in Appendix \ref{appendix:lagrangian}.)  The collective degrees of freedom include the momentum center of the wavepacket ${\bf K}_c$, as well as average hole and electron positions,
which may be written in the form
\begin{align}
\label{position_connection_wp_hole}
{\bf R}_1 \approx \vec{\nabla}_{K_c}\gamma({\bf K}_c) +{\mathcal A}^{(1)}({\bf K}_c),\\
\label{position_connection_wp_electron}
{\bf R}_2 \approx \vec{\nabla}_{K_c}\gamma({\bf K}_c) +{\mathcal A}^{(2)}({\bf K}_c),
\end{align}
where $\gamma$ is the phase of the wavepacket weight, $w({\bf K}) = |w({\bf K})|e^{-i\gamma({\bf K})}$.
Together these relations imply a constraint on the collective degrees of freedom,
\begin{equation}
\label{constraint}
{\bf R}_1-{\bf R_2}={\mathbfcal A}^{(1)}({\bf K}_c)-{\mathbfcal A}^{(2)}({\bf K}_c).
\end{equation}
We again see the intrinsic connection between the dipole moment of the exciton and the quantum geometric dipole of the exciton band.  Since the position difference is fixed in terms of ${\bf K}_c$, we in fact only have two independent (vector) degrees of freedom, which we take to be ${\bf K}_c$ and the sum position ${\bf R}_+ \equiv {\bf R}_1+ {\bf R}_2$.  In terms of these variables, as shown in Appendix \ref{appendix:lagrangian}, the Lagrangian may be written as
\begin{equation}
{\mathcal L}
= {e \over 2} \sum_{\mu=x,y} \delta A_{+,\mu} \dot{\bf K}_c \cdot \vec{\nabla}_{K_c} \left[{\mathcal A}^{(1)}_{\mu}-{\mathcal A}^{(2)}_{\mu} \right] - {e \over 2} \delta {\bf A}_- \cdot \dot{\bf R}_+ -{1 \over 2} \dot{\bf K}_c \cdot \left[{\bf R}_+
-{\mathbfcal A}^{(1)}-{\mathbfcal A}^{(2)} \right]-E_0({\bf K_c}),
\label{lagrangian_main_text}
\end{equation}
where $\delta A_{\pm} = \delta A^{(2)} \pm \delta A^{(1)}$ and $E_0({\bf K})$ is the exciton energy dispersion in the absence of $\delta A$.

Minimizing $S$ with this Lagrangian with respect to the two free parameters yields the Euler-Lagrange equations,
\begin{align}
\label{Euler_Lagrange_R}
\frac{d}{dt} \frac{\partial {\mathcal L}}{\partial \dot{{\bf R}}_+}
&= \frac{\partial {\mathcal L}}{\partial {{\bf R}}_+}, \\
\label{Euler_Lagrange_K}
\frac{d}{dt} \frac{\partial {\mathcal L}}{\partial \dot{{\bf K}}_c}
&= \frac{\partial {\mathcal L}}{\partial {{\bf K}}_c}.
\end{align}
Specializing for simplicity to the case $\delta {\mathbfcal B} = 0$, implementing Eqs. \ref{Euler_Lagrange_R} and \ref{Euler_Lagrange_K}, after some algebra, leads to
\begin{equation}
\label{Rdot}
-\dot{\bf R}_+ + e\vec{\nabla}_{K_c}\left[\delta \dot{\bf A}_+ \cdot \left({\mathbfcal A}^{(1)} - {\mathbfcal A}^{(2)} \right) \right] =
-2\vec{\nabla}_{K_c} E_0({\bf K}_c) + \dot{\bf K}_c \times \left[ \vec{\nabla}_{K_c} \times \left({\mathbfcal A}^{(2)} + {\mathbfcal A}^{(1)} \right) \right]
\end{equation}
and
\begin{equation}
\label{Kdot}
-\dot{\bf K}_c = {e}\delta \dot{\bf A}_- + {e} \vec{\nabla}_{R_+}
\left\{ \dot{\bf K}_c \cdot \vec{\nabla}_{K_c} \left[ \delta {\bf A}_+ \cdot \left({\mathbfcal A}^{(1)} - {\mathbfcal A}^{(2)} \right) \right] \right\}.
\end{equation}

Eqs. \ref{Rdot} and \ref{Kdot} are the main results of this section, and a few comments on their forms are warranted.  While the equations involve gauge-dependent quantities (${\mathbfcal A}^{(\alpha)}$), the combinations in which they appear are gauge-invariant.  The last term of Eq. \ref{Rdot} is essentially the Berry's curvature for the exciton, and enters as an anomalous velocity \cite{Marder_book}, which is known to impact the dynamics of single-particle carriers in general, as well as the dynamics of excitons \cite{Yao_2008}.  The terms involving the quantum geometric dipole ${\mathbfcal D}=\left({\mathbfcal A}^{(1)} - {\mathbfcal A}^{(2)} \right)$, by contrast, are a new element.  Their presence is possible because the exciton is a two-body object, and has no analog in the single-particle dynamics of band electrons \cite{Chang_1996}.

To better appreciate the meaning of these terms, we consider the situation in which the applied electric fields (and their corresponding vector potentials) are spatially uniform.  Writing the sum and difference of the electric fields in the two layers as ${\mathbfcal E}_{\pm}$, Eqs. \ref{Rdot} and \ref{Kdot} reduce to
\begin{align}
\label{Rdot_E}
\dot{\bf R}_+ + \dot{\bf K}_c \times \mathbf{\Omega}({\bf K}_c) &= 2\vec{\nabla}_{K_c} E_0({\bf K}_c) - e \vec{\nabla}_{K_c} \left[{\mathbfcal E}_+ \cdot {\mathbfcal D}({\bf K}_c) \right],\\
\label{Kdot_E}
\dot{\bf K}_c &= {\mathbfcal E}_-,
\end{align}
where $\mathbf{\Omega} = \vec{\nabla} \times \left({\mathbfcal A}^{(2)} + {\mathbfcal A}^{(1)} \right)$ is the effective Berry's curvature for the exciton, and ${\mathbfcal D}$ is the exciton quantum geometric dipole. Noting that the dipole moment of the exciton is ${\bf d}=e{\mathbfcal D}$, we see that the quantum geometric dipole enters the equations in a very natural way, coupling as one expects of an electric dipole moment in an electric field.  In situations where the quantum geometric dipole is constant, this will have no effect on the exciton equations of motion; however, this is generically not the case. Thus the exciton acquires a further quantum geometric correction to the velocity beyond that of the Berry's curvature.

As demonstrated in the previous section, for exciton constituents confined to Landau levels, the quantum geometric dipole is precisely linear in ${\bf K}_c$ (Eq.\ref{dipole_curv_LL}) so that this particular geometric contribution to the velocity has a simple interpretation.  Assuming the same electric field $\mathbfcal{E}$ to be present for both the hole and the electron, one finds $|{1 \over 2} \vec{\nabla}_{K_c} \left[{\mathbfcal E}_+ \cdot {\bf d} \right]|= {\mathcal E}/B \equiv v_D$, the expected drift velocity of a charge particle in crossed electric and magnetic fields.  While the exciton is a neutral object, it nevertheless couples to the electric field through its dipole moment.  The anomalous contribution to the velocity is essentially that of a frame of reference in which the electric field vanishes. \cite{Imamoglu_note} (Note that this is only possible when ${\mathcal E} < B$; since Eq. \ref{dipole_curv_LL} was derived in the strong magnetic field limit, we can assume we are in this situation.)  More generally, we will see the quantum geometric dipole varies linearly near $K_c=0$, so that its impact on the exciton equations of motion in an electric field is highly analogous to what one expects when the object is in a magnetic field.  Thus, from the perspective of dynamics, this quantum geometric phase renormalizes the effective magnetic field, even introducing an effective field where a real one is not present.  We will see an example of this in the next section.

The exciton equations of motion involve three quantities that are specific to the particular system they are describing: the dispersion $E_0({\bf K})$, the Berry's curvature ${\mathbf \Omega}$, and the quantum geometric dipole ${\mathbfcal D}$.  We next turn to computing these, first for the instructive example of a fictitious ``gapped graphene'' heterostructure, and then for magnetoexcitons in van der Waals heterostructures.

\section{Gapped Graphene in Zero Field}
\label{section:gapped_graphene}

As is apparent from the development above, the connection between an exciton dipole moment and a quantum geometric phase is not specific to systems in a magnetic field, but will occur in any system for which there is sufficient difference in the way that Bloch wavefunctions $|u_{\bf K},\alpha\rangle$ (Eq. \ref{udef}) evolve with ${\bf K}$ when the plane wave phase is removed from either the hole ($\alpha=1$) or the electron ($\alpha=2$).  It is interesting and instructive then to see how this plays out in a relatively simple setting: gapped graphene.  This can in principle be realized as two graphene sheets separated by a boron nitride spacer layer \cite{Giovannetti_2007,Hunt_2013}, with twist angles that are slightly different so that each layer can have a different induced band gap.


We thus begin with a two-body Hamiltonian of the form \hfil
\break
\begin{equation}
\label{ggHam}
H({\bf r}_1, {\bf r}_2 )=
\left[v_1\left(-i\partial_{x_1}\sigma_x -i\partial_{y_1}\sigma_y \right)  + \delta_1\sigma_z \right] \otimes \mathbbm{1}
+
\mathbbm{1} \otimes \left[v_2\left(-i\partial_{x_2}\sigma_x -i\partial_{y_2}\sigma_y \right)  -\delta_2\sigma_z
\right] + v({\bf r}_1-{\bf r}_2)
\end{equation}
\break
where ${\bf r}_1$ is the hole position, ${\bf r}_2$ the electron position, $v$ is the attractive electron-hole interaction, and the matrices act on the wavefunction amplitudes for the graphene $A$ and $B$ sites.  Note in writing this, we have implemented a particle-hole transformation for the hole degree of freedom as described above, $H^{(hole)}_{1-body}({\bf k}) = -H^{(electron)}_{1-body}(-{\bf k})$, accounting for the differing sign signatures for the gap parameters, $\delta_1$ and $\delta_2$.  If both layers are composed of graphene then the Fermi velocities, $v_1$ and $v_2$, should be the same, but we allow them here to be different for further generality.  Note that this model also describes the low energy physics of other materials, such as topological insulator surface states \cite{Garate_2011,Brey_2014}, with spinor amplitudes referring to spin states, or TMD materials \cite{Xiao_2012}.

We wish to find minimum energy states of Eq. \ref{ggHam}, subject to the constraint that they involve only positive energy states of the single-body Hamiltonians for the electron and hole. To accomplish this, we change variables, writing
\begin{eqnarray}
{\bf R} & =& \beta_1 {\bf r}_1 +\beta_2 {\bf r}_2, \\
{\bf r} &=& {\bf r}_2 -{\bf r}_1,
\end{eqnarray}
with the constraint $\beta_1 +\beta_2 =1$.  The values of $\beta_{1,2}$ are for now arbitrary and will be chosen for convenience in what follows.  Since the total  ${\bf K}$
is a constant of motion, the exciton wave function can be written as
\begin{equation}
\label{separate_wf}
\Phi _{{\bf K}} = e ^{i{\bf K}\cdot {\bf R}} u_{\bf K}  ({\bf r}).
\end{equation}
We expand the relative wavefunction $u_{\bf K}({\bf r})$ in the form
\begin{equation}
\label{ukdef}
u_{{\bf K}} ({\bf r})=\frac{1}{\sqrt{L_xL_y}}\sum _{\bf q}  \, C_{\bf q}({\bf K}) \psi _e ({\bf q},{\bf K}) \psi_h ({\bf q},{\bf K}) \,
e^{i {\bf q}{\bf r}},
\end{equation}
where $L_xL_y$ is the sample area.
The (positive energy) single particle wavefunctions appearing in this expression have the forms

\begin {equation}
\label{single_particle_states}
\psi _h ({\bf q},{\bf K}) =  \left (  \begin{array} {c} \cos {\frac {\theta _1} 2}  \\  \sin {\frac {\theta _1} 2} e^{i \phi_1}
\end {array} \right ),
\quad
\psi _e ({\bf q},{\bf K}) =  \left (  \begin{array} {c} \sin {\frac {\theta _2} 2}  \\  \cos {\frac {\theta _2} 2} e^{i \phi_2}
\end {array} \right ),
\end{equation}
with energies $E_e=\sqrt { \delta _2 ^2 + v_2 ^2 ( \beta_2 {\bf K} +{\bf q} ) ^2 } $ and $E_h=\sqrt { \delta _1 ^2 + v_1 ^2 ( \beta_1 {\bf K} -{\bf q} ) ^2 }$, and the angular parameters are given by
$
\theta_i =\tan ^{-1}
\frac {v_i|\beta_i {\bf K}+(-1)^i{\bf q}|}{\delta _i} \, \, \, {\rm and} \, \,  \phi_i =\tan ^{-1} \frac {\beta_i K_y + (-1)^i q_y }{\beta_i K_x + (-1)^i q_x}.
$

Minimizing $\langle \Phi_{\bf K} | H | \Phi_{\bf K} \rangle$ subject to the constraint that the wavefunction is normalized, one finds the coefficients $C_{\bf q}({\bf K})$ must obey an eigenvalue equation. For this particular case, it turns out to be useful to absorb a phase into these coefficients, writing $\bar{\chi}({\bf q},{\bf K})=e^{i\phi_2({\bf q},{\bf K})}C_{\bf q}({\bf K})$.  In terms of this the eigenvalue equation is
\begin{equation}
\left[ E_e ({\bf q},{\bf K}) +E_h ({\bf q},{\bf K}) \right] \bar{\chi}({\bf q},{\bf K})
+\frac 1 {L_xL_y} \sum _{{\bf q}'} \bar{V}({\bf q},{\bf q}',{\bf K})\bar{\chi}({\bf q}',{\bf K})  =E({\bf K}) \bar{\chi}({\bf q},{\bf K})
\end{equation}
where the projected interaction has the form
\begin{equation}
\bar{V}({\bf q},{\bf q}',{\bf K})= 
\tilde{v}({\bf q}-{\bf q}')
\langle\psi _e ({\bf q},{\bf K})|\psi _e ({\bf q}',{\bf K})\rangle  \langle\psi _h ({\bf q}',{\bf K})|\psi _h ({\bf q},{\bf K}) \rangle e^{i\left[\phi_2({\bf q},{\bf K})-\phi_2({\bf q}',{\bf K})\right]}
\end{equation}
with $\tilde{v}$ the Fourier transform of the particle-hole interaction.  Note that because we are working with spatially separated layers, we have neglected any effects of exchange interactions.

To make further progress, we consider excitons of relatively low total momentum $K$, and moreover assume $\bar{\chi}_{\bf q}({\bf K})$ is only significant for small values of $q$ for such excitons, which we find self-consistently to be the case.  With considerable algebra, one may show
\begin{eqnarray}
\label{Vbar}
{\bar V} ({\bf q},{\bf q}',{\bf K}) & \approx & \tilde{v}({\bf q}-{\bf q}')
( 1 \! - \! \left (  \frac {v_1 ^2}{ 8 \delta _1 ^2} \!+\! \frac {v_2 ^2}{ 8 \delta _2 ^2} \right )
 ({\bf q} \! - \!  {\bf q}') ^2 \nonumber \\
&+& i
\left (  \frac {v_1 ^2}{ 4 \delta _1 ^2} \!+\! \frac {v_2 ^2}{ 4 \delta _2 ^2} \right )
\hat z \cdot ({\bf q} \times {\bf q}')+
i \left (  \frac {v_1 ^2}{ 4 \delta _1 ^2} \beta_2 \! - \! \frac {v_2 ^2}{ 4 \delta _2 ^2} \beta_1  \right )
\hat z \cdot ({\bf K} \times ({\bf q} \! - \!{\bf q}'))  ).
\end{eqnarray}
In the same approximation,
\begin{equation}
\label{single_particle_energies}
E_e ({\bf q},{\bf K}) +E_h ({\bf q},{\bf K}) \approx \delta_1 +\delta_2 + \frac {q^2}2 (\frac {v_1^2}{\delta _1}+ \frac {v_2 ^2}{\delta _2}) +
 {\bf K} \cdot {\bf q} ( \beta_2 \frac {v_1^2}{\delta _1} -\beta_1 \frac {v_2 ^2}{\delta _2})
+\frac {K^2} 2 (\beta_2 ^2  \frac {v_1^2}{\delta _1} +\beta_1 ^2 \frac {v_2 ^2}{\delta _2}).
\end{equation}

Eqs. \ref{Vbar} and \ref{single_particle_energies} admit a special case where explicit solutions can essentially be written down: $\delta_1=\delta_2 \equiv \delta$.  In this case one may choose
$\beta_i$ to cancel terms linear in ${\bf K}$ in these equations, and $K^2$ enters only as a constant energy shift in the eigenvalue equation.  The energy disperses quadratically with total momentum, with a curvature independent of the internal state of the exciton.  Moreover, the amplitude $\bar {\chi}$ solving the eigenvalue equation does not depend on ${\bf K}$.  In the remainder of this section, we focus on this special case.

With the form of the wavefunction we have used in Eq. \ref{separate_wf}, we can consider the Berry's connection in a more standard approach \cite{Yao_2008,Kuga_2008,Qiu:2015aa,Wu_2017,Kwan_2020}, in which we identify the plane wave part of the wavefunction with the center coordinate, in this section written as ${\bf R}$.  Thus one takes
\begin{equation}
{\mathbfcal A}( {\bf K})=i < u _{\bf K}| \vec{\nabla} _{\bf K}| u_{\bf K}>,
\end{equation}
with $|u_{\bf K}\rangle$ the ket corresponding to the state in Eq. \ref{ukdef}.
Because $\bar{\chi}$ is independent of ${\bf K}$, using Eqs. \ref{single_particle_states} one can obtain a rather simple formal expression for ${\mathbfcal A}$,
\begin{equation}
\label{standard_berrys_connection}
{\mathbfcal A}( {\bf K})=
\sum_{\bf q} |\bar{\chi}(q)|^2\left[ \sin^2\frac{\theta_2}{2}\vec{\nabla}\phi_2 - \sin^2\frac{\theta_1}{2}\vec{\nabla}\phi_1 \right].
\end{equation}

This quantity, however, is not quite what we want: we need the appropriate Berry's connections that can be formed into the quantum geometric dipole. To do so we consider states of the form
\begin{align}
u_{\bf K}^{(\alpha)}({\bf r}_1,{\bf r}_2) & =e^{-i{\bf K} \cdot \left[ \alpha {\bf r}_1 + (1-\alpha) {\bf r}_2 \right]}\Phi_{\bf K} \nonumber\\
&= e^{i{\bf K} \cdot \left[ (\beta_1-\alpha){\bf r}_1 + (\beta_2 + \alpha -1){\bf r}_2 \right]}u_{\bf K}({\bf r}),
\label{uk_to_ukalpha}
\end{align}
which are the real space representations of the kets defined in Eq. \ref{udef}.  Using $\beta_1+\beta_2=1$ we then have
\begin{equation}
u_{\bf K}^{(\alpha=0)}=e^{-i\beta_1 {\bf K} \cdot {\bf r}} u_{\bf K}({\bf r}), \quad
u_{\bf K}^{(\alpha=1)}=e^{i\beta_2 {\bf K} \cdot {\bf r}} u_{\bf K}({\bf r}).
\end{equation}
Writing
\begin{equation}
\label{ukguts}
U({\bf q}, {\bf K}) \equiv C_{\bf q}({\bf K}) \psi _e ({\bf q},{\bf K}) \psi_h ({\bf q},{\bf K}),
\end{equation}
we obtain
\begin{align}
\label{Acurv}
{\mathbfcal A}^{(1)} &= i\langle u_{\bf K},\alpha=1| \vec{\nabla}_{\bf K} | u_{\bf K},\alpha=1\rangle\nonumber\\
&=\frac{1}{L_xL_y}\sum_{{\bf q},{\bf q}'} \int d^2 r U^{\dag}({\bf q}, {\bf K})e^{i\left[\beta_1 {\bf K} - {\bf q}\right] \cdot {\bf r}} \vec{\nabla}_{\bf K} e^{-i\left[\beta_1 {\bf K} - {\bf q}'\right] \cdot {\bf r}}U({\bf q}', {\bf K}) \nonumber\\
&={\mathbfcal A}({\bf K}) + i\frac{1}{L_xL_y}\sum_{{\bf q},{\bf q}'}U^{\dag}({\bf q}, {\bf K})U({\bf q}', {\bf K}) \int d^2 r e^{i\left[\beta_1 {\bf K} - {\bf q}\right] \cdot {\bf r}} \vec{\nabla}_{\bf K} e^{-i\left[\beta_1 {\bf K} - {\bf q}'\right] \cdot {\bf r}} \nonumber\\
&={\mathbfcal A}({\bf K}) + i{\beta_1}\sum_{{\bf q}}U^{\dag}({\bf q}, {\bf K})\vec{\nabla}_{\bf q}U({\bf q}, {\bf K}).
\end{align}
A similar manipulation yields an analogous expression for ${\mathbfcal A}^{(2)}$, and using $\beta_1+\beta_2=1$, one finds
\begin{equation}
\label{gapped_graphene_curvature}
{\mathbfcal A}^{(1)}-{\mathbfcal A}^{(2)} = i\sum_{{\bf q}}U^{\dag}({\bf q}, {\bf K})\vec{\nabla}_{\bf q}U({\bf q}, {\bf K}).
\end{equation}

For $\delta _1=\delta _2= \delta$, taking $\beta_1= \frac {v_1^2} {v_1^2+v_2^2}$ and $\beta_2 =\frac {v_2^2}{v_1^2 + v_2^2}$,
we reach an expression analogous to Eq.\ref{standard_berrys_connection},
\begin{eqnarray}
\label{Ugg}
\sum_{\bf q} U^{\dag}({\bf q}, {\bf K})\vec{\nabla}_{\bf q}U({\bf q}, {\bf K})
= -i \sum _{q}   \left\{ i{\bar \chi }^*({\bf q}){ \nabla} _{\bf q}  {\bar \chi }({\bf q})
+  |{\bar \chi }({\bf q})|^2  \sin ^2 {\frac {\theta _2} 2} \, \,  { \nabla} _{\bf q}  { \phi_2 ({\bf q},{\bf K})}
- |{\bar \chi }({\bf q})|^2 \sin ^2 {\frac {\theta _1} 2}\, \,
 { \nabla} _{\bf q}  { \phi_1 ({\bf q},{\bf K})} \right\}. \nonumber\\
\end{eqnarray}

To make further progress,
it is necessary to have an explicit expression for the amplitude $\bar \chi$. In order to obtain qualitative results, we model the relative exciton wavefunction as a 1S state of the hydrogen atom, with an effective mass $\mu = m^* /2$ where $m^ *=\frac {2 \delta}{v_1^2 +v_2^2}$.  Thus we take
\begin{equation}
\varphi _{1S}= \frac {2 \sqrt{2}}{a _0 \sqrt{\pi}} e ^{-2 r /a_0} \, \, \, \, \Rightarrow \, \, \,
\bar \chi (q)= \sqrt{2 \pi} \frac {a_0} {(1+ \frac {q^2 a_0^2} 4)^{\frac 3 2}}
\end{equation}
with $a_0=\frac {\hbar ^2}{e^2 \mu}$.
(Comparison with results from Ref. \onlinecite{Zhou_2015}
indicate that this expression is quite reasonable.)  Note that in this model the first term on
the right hand side of Eq. \ref{Ugg} vanishes, so that the form of the quantum geometric dipole comes out rather
similar to the standard Berry's connection, Eq. \ref{standard_berrys_connection}, in the gauge we use.
After some algebra one can write an explicit integral form for Eq. \ref{gapped_graphene_curvature},


\begin{eqnarray}
&  {\mathbfcal A} ^{(2)} ({\bf K}) -{\mathbfcal A} ^{(1)} ({\bf K})
 =  \frac {a_0} {2 \pi} \left( \hat{z} \times \hat{K} \right)\int dk d\zeta
\frac {\cos \zeta} 2 \nonumber \\
& \times   \left[
\left (\frac {1}{\sqrt{1+\beta_2 k^2}}-1 \right )
\left (
\frac 1 {(1+\frac 1 4 (k^2+\beta_2^2K^2-2\beta_2 k K \cos \zeta))^3} \right)
-
\left (\frac {1}{\sqrt{1+\beta_1 k^2}}-1 \right )
\left (\frac 1 {(1+\frac 1 4 (k^2+\beta_1^2K^2+2\beta_1 k K \cos \zeta))^3} \right ) \right].
\end{eqnarray}

\begin{figure}
  \centering
  \includegraphics[width=0.6\textwidth,trim= 0 0 0 0,clip]{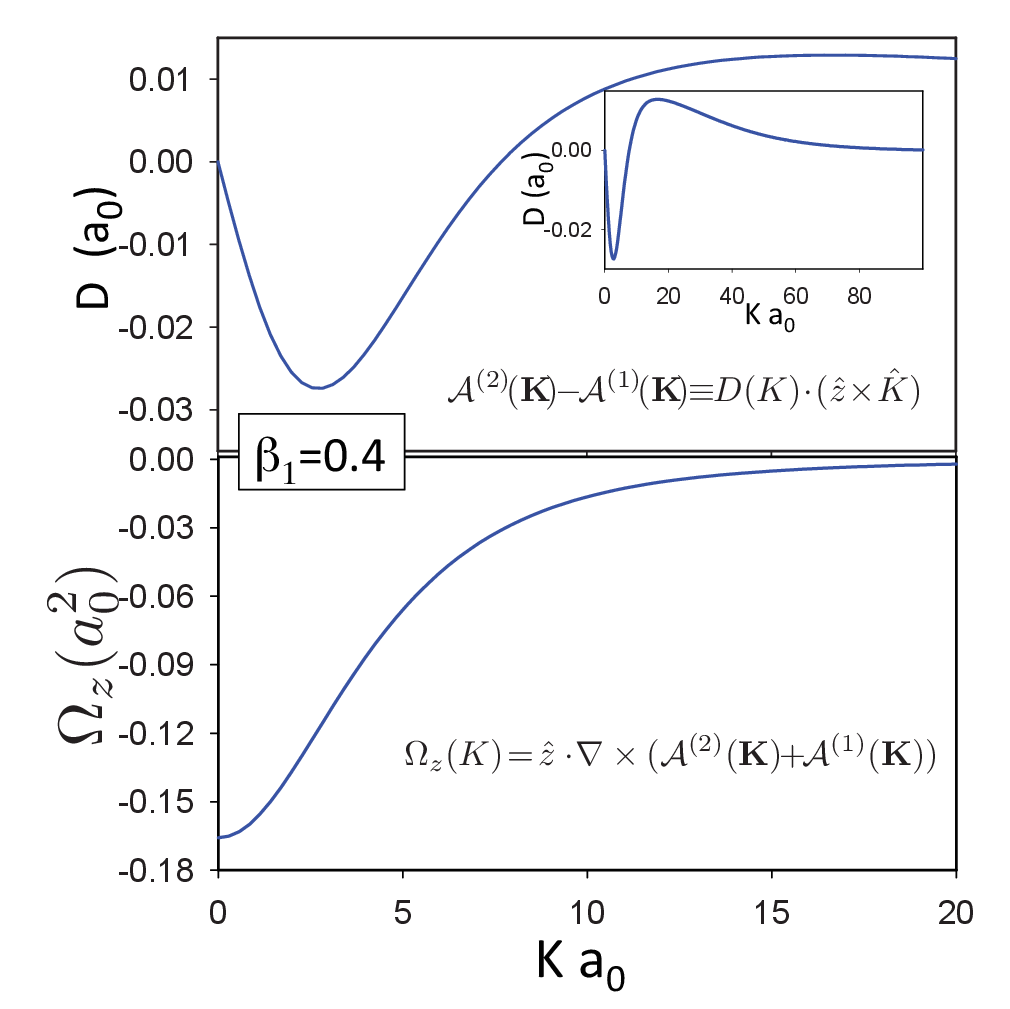}
  \caption{Upper panel: Magnitude of quantum geometric dipole ${\mathbfcal D} =  {\mathbfcal A}^{(2)} - {\mathbfcal A}^{(1)}$ for two different gapped graphene layers, with $\beta_1 = 0.4$ (see text).  Lower panel: Berry's curvature for the same system.}
  \label{gapped_G_fig}
\end{figure}

Fig. \ref{gapped_G_fig} shows an illustration of the quantum geometric dipole and Berry's curvature for a representative set of parameters.    Note that for small values of $K$ the quantum geometric dipole is linear in $K$, and is directed perpendicular to ${\bf K}$.  This form is exactly as was found for excitons in a strong magnetic field, and, according to the considerations of Section \ref{semiclassical}, indicates that we expect the exciton to drift perpendicular to an applied electric field, precisely as if the object were immersed in a perpendicular magnetic field -- although, in this example, no real field is present.

Our formulation of the quantum geometric dipole and Berry's curvature also have interesting behaviors when $\delta_1=\delta_2$ and $v_1=v_2$; i.e., when the layers are identical.  {In this case $\beta_1=\beta_2$, and one easily sees that the quantum geometric dipole vanishes.}  This situation applies if we consider the quantum geometric dipole of an exciton in a {\it single} layer gapped graphene system: excitons in this simpler system do not carry a dipole moment.  Physically, the definition of exciton Berry's curvature we use allows for the coupling to electric fields that differ for the two constituents (cf. Eq. \ref{Kdot_E}), for which we expect to have a non-vanishing response. Note this situation is realizable in principle for interlayer excitons, but could not occur (at least for electric fields that vary slowly in space) in a single layer system.  We thus see that heterostructures can yield interesting quantum geometric behavior which is absent in single layer systems.

\newpage

\section{Magnetoexcitons in Heterostructures}
\label{section:Heterostructures}
We now turn to investigating properties of interlayer excitons in graphene/TMD heterojunctions, as well as heterojunctions of different single-layer TMD materials. The method we follow parallels much of the development in Section \ref{section:LandauLevels}:
we adopt trial wavefunctions for the exciton generically that are analogous to Eq. \ref{exciton_wf1},
\begin{equation}
\Phi_{\bf K} = \sum_{\bf{n}\bf{q}} C_{{\bf n}}({\bf K}) e^{i{\bf q}\cdot {\bf R}}e^{iK_x^{(0)}q_y \ell^2} \vec{\phi}^{(h)}_{n_h,{\bf q},\sigma_h,\tau_h}({\bf r}_1) \otimes \vec{\phi}^{(e)}_{n_e,{\bf K^{(0)}}-{\bf q},\sigma_e,\tau_e}({\bf r}_2),
\end{equation}
where the full momentum ${\bf K}={\bf K}^{(0)} -\hat{z} \times {\bf R}/\ell^2$, has been broken up into a part that lies in the first Brillouin zone, ${\bf K}^{(0)}$, and a reciprocal lattice vector, $-\hat{z} \times {\bf R}/\ell^2$,
where ${\bf R}$ is a direct lattice vector, as discussed in Sec. \ref{section:LandauLevels}.
The vector ${\bf n} \equiv (n_1,n_2)$ represents the set of possible Landau level indices the hole ($n_1$) and the electron ($n_2$) may have (i.e., the set of levels which are not Pauli-blocked by other carriers.)  The wavefunctions $\vec{\phi}^{(h)}_{n_h,{\bf q},\sigma_h,\tau_h}({\bf r}_1)$ and $\vec{\phi}^{(e)}_{n_e,{\bf K^{(0)}}-{\bf q},\sigma_e,\tau_e}({\bf r}_2)$ are eigenstates of the single-particle Hamiltonians for the hole and electron, respectively, with the hole of spin $\sigma_h$ residing in valley $\tau_h$, and the electron of spin $\sigma_e$ residing in valley $\tau_e$.

To find the exciton energies and wavefunctions, we minimize the energy $\langle \Phi_{\bf K} | H | \Phi_{\bf K}\rangle$ with respect to the coefficients $C_{{\bf n}}({\bf K})$, subject to the constraint that $|\Phi_{\bf K}\rangle$ is normalized.  The procedure is conceptually analogous to what was described in Section \ref{section:LandauLevels}, although the details are considerably more involved. At the end of the analysis one finds an eigenvalue equation for the exciton energies and wavefunctions that can be solved numerically.  From the latter, the quantum geometric dipole and Berry's curvature may be ascertained.

We next turn to the results of these calculations.
Readers interested in further details of the eigenvalue equation, as well as in some discussion of our numerical approaches, may find these in Appendix \ref{appendix:exciton_equations}.

\subsection{Graphene on TMD}

We first discuss the behavior of an exciton in which the electron resides in a graphene layer while the hole resides in a TMD layer.  For concreteness, we take the TMD to be MoS$_2$, although we do not expect qualitatively different results for other TMDs.  Both materials are honeycomb lattices, but the lattice constant for MoS$_2$ is considerably smaller than that of graphene (0.318nm, vs. 0.246nm) \cite{Molina_2011,Castro_Neto_2009}.  In this situation the tunneling between the two layers is likely to be quite small, but we assume it is sufficient that luminescence from electrons and holes in different layers recombining may be detected.  Note that in the absence of a magnetic field, radiation from electrons and holes recombining from these two different materials is challenging to detect due to the semimetallic nature of (ungapped) graphene \cite{Yang_2018,Lorchat_2020,Subramanian_2020}; the gap provided by the field however renders the graphene insulating in the situations we consider here.

Electrons and holes may be introduced into the layers through tunneling, direct excitation by light, or by simply by injecting them into the individual layers.  Note that because of the large lattice mismatch, one does not expect direct excitation into exciton states by light absorption: since light absorption conserves momentum, one or both carriers will be far from a $K$ or $K'$ when the particle-hole pair is created, so that they will be effectively unbound.  In TMD heterostructures it is known that injected electrons and holes rapidly relax into exciton states, which are then long-lived; we assume the situation is similar for the structure we consider here \cite{Jauregui_2019}.

Because the field-free graphene spectrum is gapless, it is important to introduce the magnetic field.  In this situation (see Appendix \ref{appendix:exciton_equations}) the Landau level spectrum of graphene provides gaps into which the Fermi energy can be placed, allowing us to consider the situation in which an extra particle in the graphene and a hole in the TMD can be usefully analyzed as a two-body problem.
(Note that in general the center of the Landau level gap below which states are occupied in the graphene layer does not align with the center of the TMD energy gap; however in principle these can be aligned by a perpendicular electric field.  For simplicity we assume here, and in the examples that follow, that these gap centers are indeed energetically aligned.)
The situation in this combination of materials is particularly interesting in that, in zero field, the electronic structure of graphene near (but not precisely at) the Dirac point carries no Berry's curvature, whereas TMDs near their valley maxima have strong curvatures that depend on the carrier spin \cite{Xiao_2012}.  Such interlayer excitons thus allow us to explore a situation in which an curvature effects due to band structure comes in through only one of the two constituents.

Fig. \ref{G_MoS2_K_results} illustrates representative results for a magnetoexciton in a graphene-MoS$_2$ heterostructure.  In this example there is a hole with spin such that it lies in the highest energy MoS$_2$ valence band of the $K'$ valley, and an electron in the $K'$ valley of graphene.  Panel (a) shows the energy dispersions for the three lowest-lying excitons of this structure.  One surprising notable feature is a minimum at momentum $|{\bf K}| \ell \approx 0.45/\ell$.  Because of the long-wavelength treatments we adopt for our single-particle Hamiltonians, the exciton ground state is rotationally degenerate.  An interesting possibility to consider in this context is what would happen should enough excitons be created at low temperature that they Bose-condense.  The resulting state will spontaneously break not just a U(1) symmetry associated with exciton number \cite{Paquet_1985,Fertig_1989}, but a second U(1) symmetry associated with rotations.  Because of the quantum geometric dipole, this would be detectable as a macroscopic dipole moment developing in the system.

Panels (b) and (c) illustrate the {\it deviation} of the quantum geometric dipole $\delta {\mathbfcal D}$ from the single Landau level results presented in Section \ref{section:LandauLevels}, ${\bf K} \times \hat{z}\ell^2$.  This computed correction is oriented along ${\hat K} \times \hat{z}$ and adds to the single Landau level result, {\it enhancing} the exciton dipole moment.  One interesting consequence of this is that, in a uniform electric field, according to the semiclassical equations of Section \ref{semiclassical}, excitons will drift faster than naively expected.  The correction is considerable: for the lowest level, our computed quantum geometric dipole at 10 T is roughly 70\% larger than the single Landau level result, which is quantitative only at much stronger fields.  It is also notable that the slopes of the quantum geometric dipole corrections for higher energy excitons are smaller than those of lower energy, and can even change sign with increasing energy at fixed ${\bf K}$ (not shown).  This impacts the speed at which they drift in an electric field.
We discuss the impact of these quantum geometric dipole enhancements on semiclassical motion of the excitons in further detail below.

Panels (d), (e), and (f) illustrate the Berry's curvature for the lowest three levels.
In general, we find
some structure near ${\bf K}=0$ for the exciton levels, but a tendency to vanish for larger $|{\bf K}|$ (not shown in figures).   The distinctions among the different exciton levels become most noticeable at small momentum, where for example the Berry's curvature for the lowest level is strongly peaked, whereas for the next two levels the structure is relatively subtle.
These differences impact the semiclassical motion of the excitons in an electric field, which we will discuss in the next section.

\begin{figure}
  \centering
\includegraphics[width=0.5\textwidth,trim= 0 0 0 0,clip]{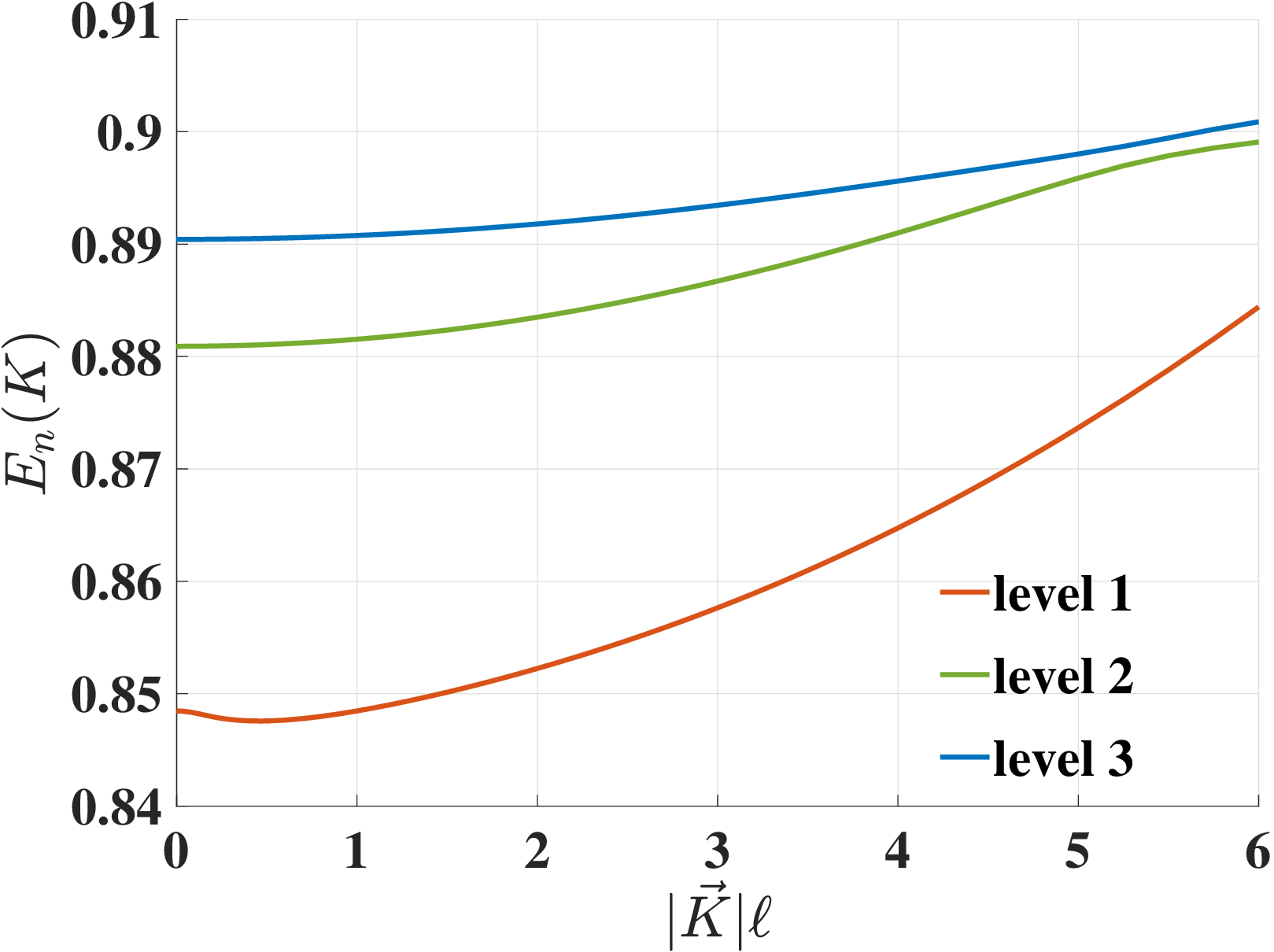}
  \caption{Energy dispersion for a graphene/TMD heterostructure. Parameters are as in Fig. \ref{G_MoS2_K_results}, except hole is in the $K$ valley. Energy is in units of eV.}
  \label{G_MoS2_Kpr_results}
\end{figure}

We also present results for an exciton in which the hole resides in the $K$ valley in Fig. \ref{G_MoS2_Kpr_results}.  (Because the wavefunctions in graphene are so similar in the two valleys, the results are independent of which valley the electron is in.)  The main difference between the single particle hole states is the absence of an $n=0$ level in the $K$, whereas one is present in the $K'$ valley.  Although the states formally reside in linear combinations of the two orbitals, over the relevant range of energies for the parameters of this system, the hole states reside largely on opposite orbitals for each of the two valleys. Moreover, the set of states associated with each of these orbitals includes a full set of harmonic oscillator wavefunctions for each value of single-particle momentum $k$, so that the states are essentially the same as those of scalar particles. Thus the main effect of the single-particle Berry's curvatures is to shift the energy dispersions, as is apparent in comparing the dispersions in Fig. \ref{G_MoS2_Kpr_results} to those of Fig. \ref{G_MoS2_K_results} \cite{Srivastava_2015, Zhou_2015}.  The shapes of the dispersions are only subtly different.  The quantum geometric dipole and Berry's curvature (not shown) are nearly identical to those of Fig. \ref{G_MoS2_Kpr_results} as well, with only subtle quantitative differences.   Thus we expect to obtain similar dynamical behavior for excitons in both valleys, although the thermal populations of the $K$ valleys should be smaller than those of the $K'$ valleys due to their higher energies.

\subsection{Gapped Graphene on Gapless Graphene}

To illustrate the effects of a more involved wavefunction structure, we briefly discuss numerical results for excitons in heterostructures composed of a gapped graphene layer placed upon an ungapped graphene layer. The former is set to have an energy gap of 111.4meV, which is the same energy difference as that between the first excited Landau level in ungapped graphene and its zero energy state for the magnetic field we adopt (10 T). Although somewhat artificial, this example allows us to see how the exciton behaviors change across valleys when the effect of the spinor structure of the wave function comes into play, in contrast to the situation
for the TMD materials discussed above, where for the relevant range of energies the single particle wave functions were essentially those of scalar particles. Note that gapped graphene can in principle be prepared by aligning it on an hBN substrate, although the resulting gap is considerably smaller than what we use here.

%

\begin{figure}[!tbp]
  \centering
  \subfloat[Energy dispersion for hole in $K$ valley, {in units of eV.}]
  {\includegraphics[width=0.45\textwidth]{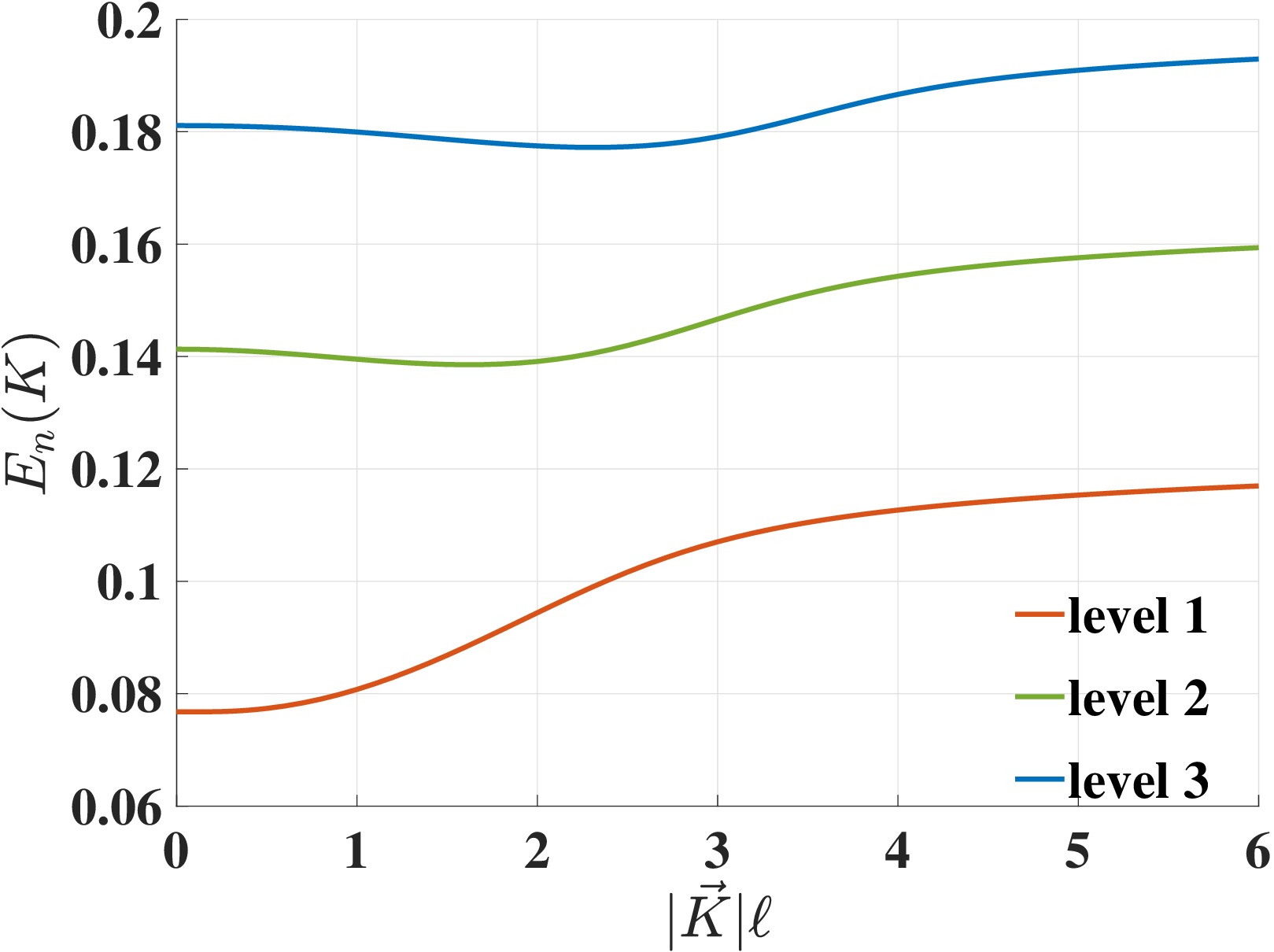}
  \label{fig:gappedG_dispersion_K_valley}}
  \hfill
  \subfloat[Energy dispersion for hole in $K'$ valley, {in units of eV.}]
  {\includegraphics[width=0.45\textwidth]{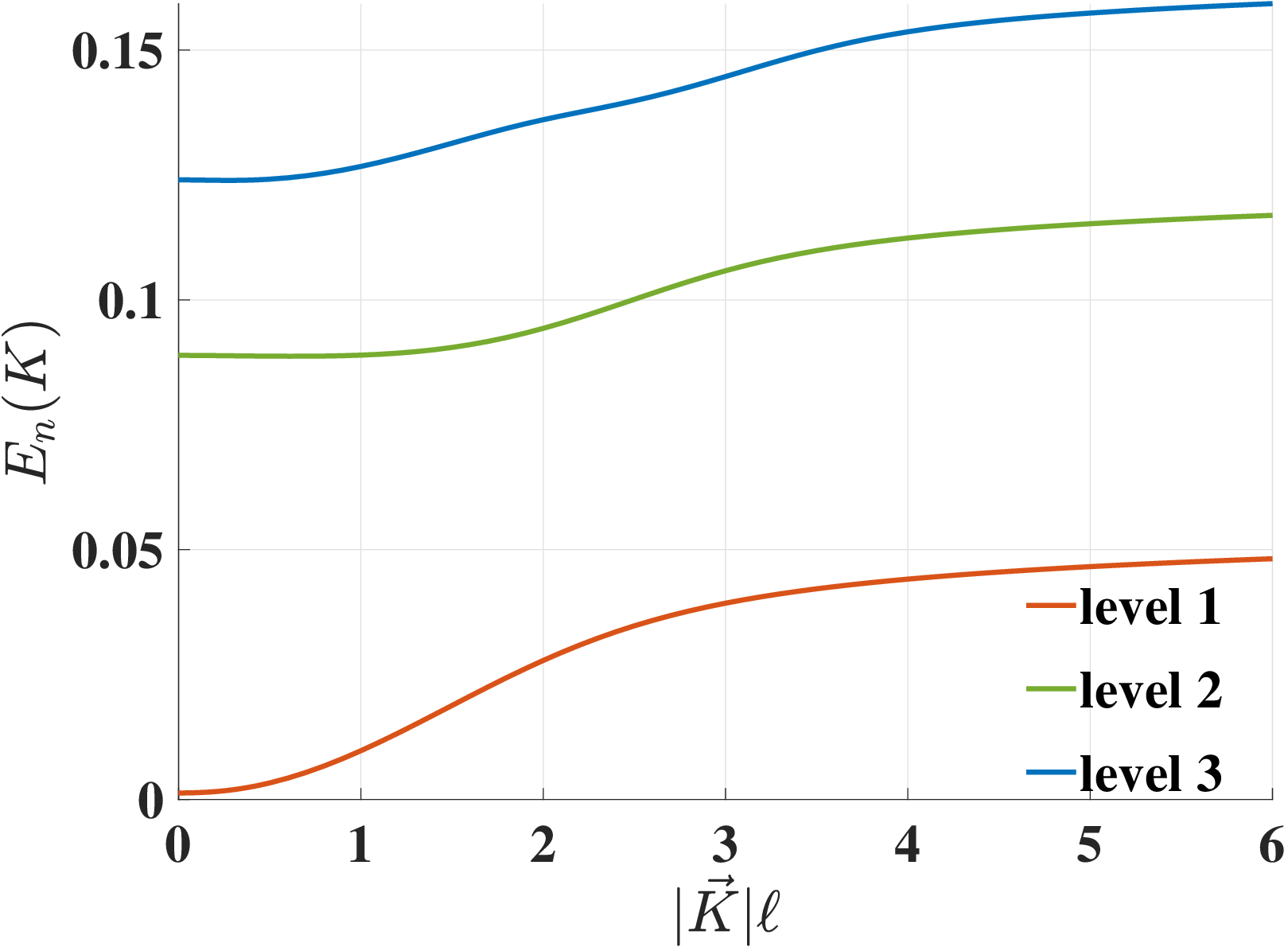}
  \label{fig:gappedG_dispersion_K'_valley}}
  \caption{Energy dispersion for exciton in a graphene/gapped graphene heterostructure, with electron and hole in different valleys.  Hole is assumed to be in the gapped graphene, and gap parameter is chosen as 111.4meV.
  Magnetic field is 10 T, interlayer separation 0.71nm.  Dielectric constant for Coulomb interaction taken to be 4.  Calculation retains states with Landau level index $|n| \le 30$ for both carriers.    \label{fig:gappedG_dispersion}
    }
\end{figure}

Fig. \ref{fig:gappedG_dispersion} presents the energy dispersion for the two valleys. Panel (a) shows results for the hole residing in a $K$ valley, and panel (b) shows them for the hole in the $K'$ valley. In addition to the overall energy shift, one sees significant differences in the shapes of the energy dispersion curves beginning in the second exciton level.  This behavior is in marked contrast to what we found for the graphene/TMD structures above.
\begin{figure}[!tbp]
  \centering
  \subfloat[Exciton dipole corrections for hole in $K$ valley.]
  {\includegraphics[width=0.49\textwidth]{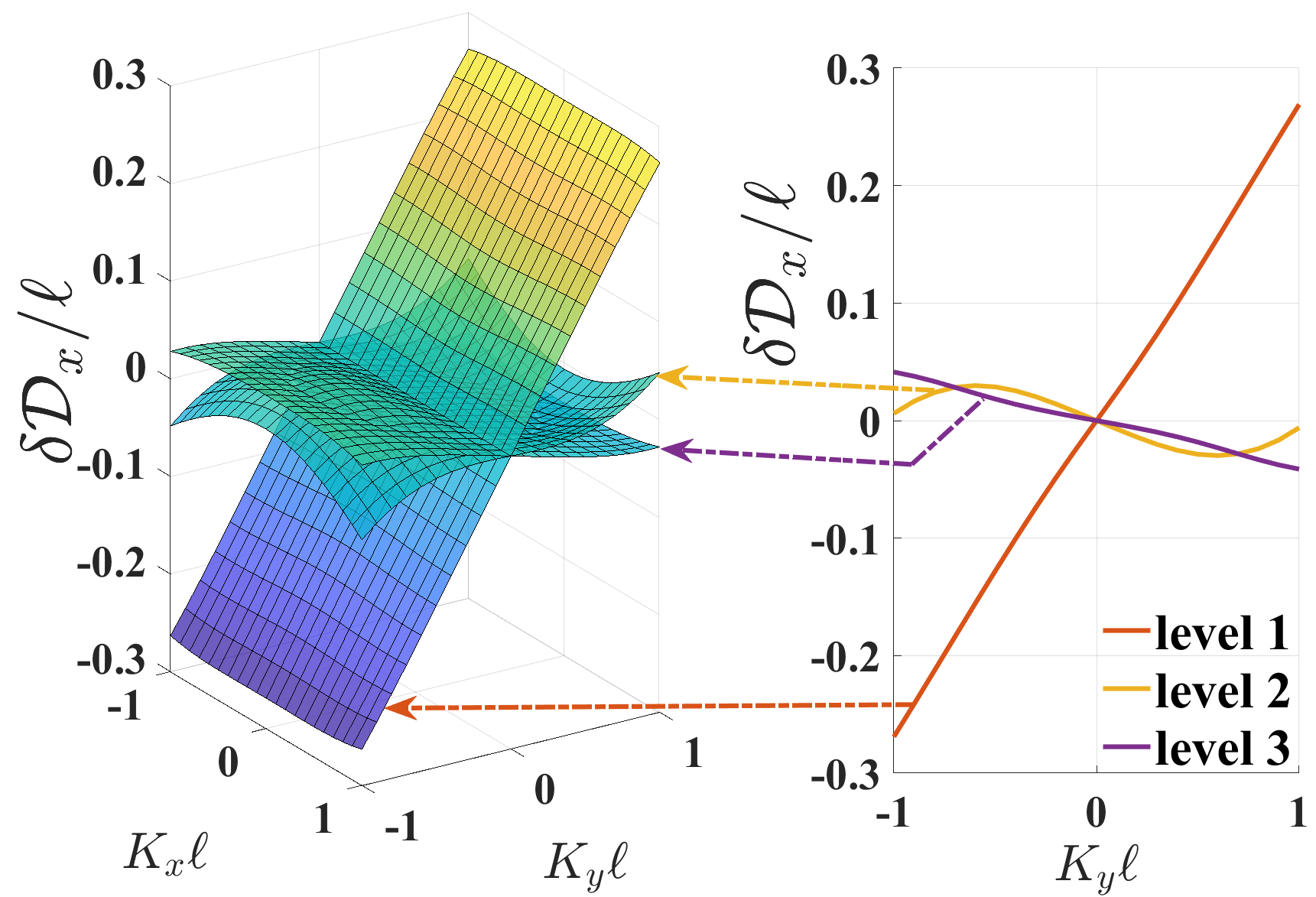}
  \label{fig:gappedG_dipole_K_valley}}
  \hfill
  \subfloat[Exciton dipole corrections for hole in $K'$ valley.]
  {\includegraphics[width=0.49\textwidth]{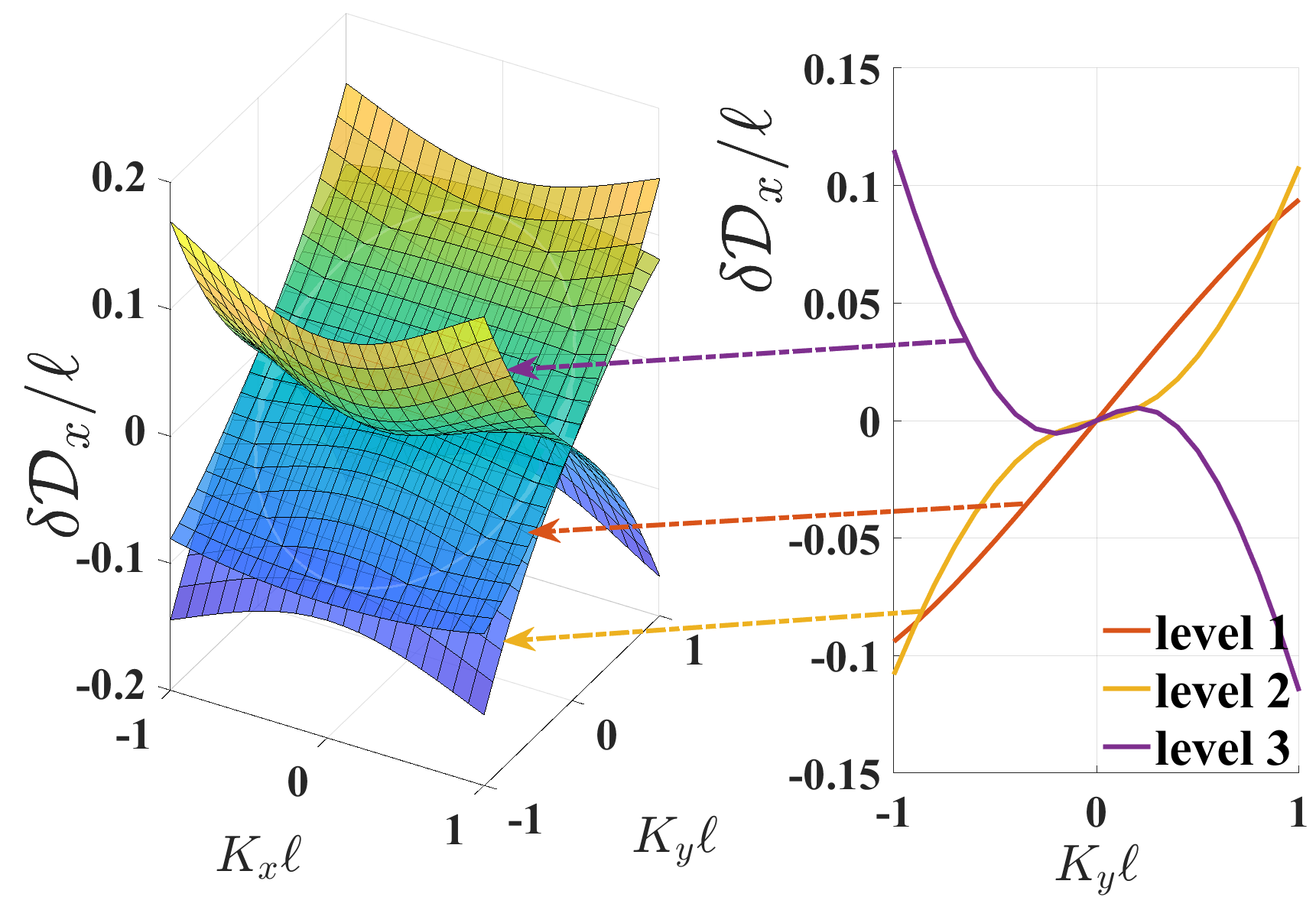}
  \label{fig:gappedG_dipole_K'_valley}}
  \caption{Correction to quantum geometric dipole $\delta {\mathbfcal D}$ beyond the single Landau level result ($\vec{K}\times\hat{z}\ell^2$) for a graphene/gapped graphene heterostructure, with hole in the gapped graphene, for three lowest exciton levels.  Parameters are those of Fig.\ref{fig:gappedG_dispersion}. Left plots show $\hat{x}$ component of  $\delta {\mathbfcal D}$ vs. momentum ${\bf K}$.  Right plots illustrate $|\delta {\mathbfcal D}| \equiv \delta {\mathcal D}_x$ assuming momentum points in $\hat{y}$ direction.
   This correction enhances the single Landau level result when its values is positive.
   \label{fig:gappedG_Dipole_x}
   }
\end{figure}

Fig. \ref{fig:gappedG_Dipole_x} illustrates the quantum geometric dipole correction for the hole in the two different valleys. In both cases the resulting contribution to the drift velocity from this changes sign at small momentum. The overall scale of the correction for lowest level for the hole in the $K$ valley is approximately twice that when it is in the $K'$ valley,  an effect again arising from the differing wavefunction structures.  It is interesting to note that, if both layers have the same gap (not shown), one finds
that the quantum geometric dipole correction vanishes.  From the perspective of the quantum geometric dipole, this again emphasizes that heterostructures offer a more interesting environment for excitons than is the case for structures involving a single material.

\begin{figure}[!tbp]
  \centering
  \subfloat[Exciton Berry's curvature for hole in $K$ valley.]
  {\includegraphics[width=0.5\textwidth]{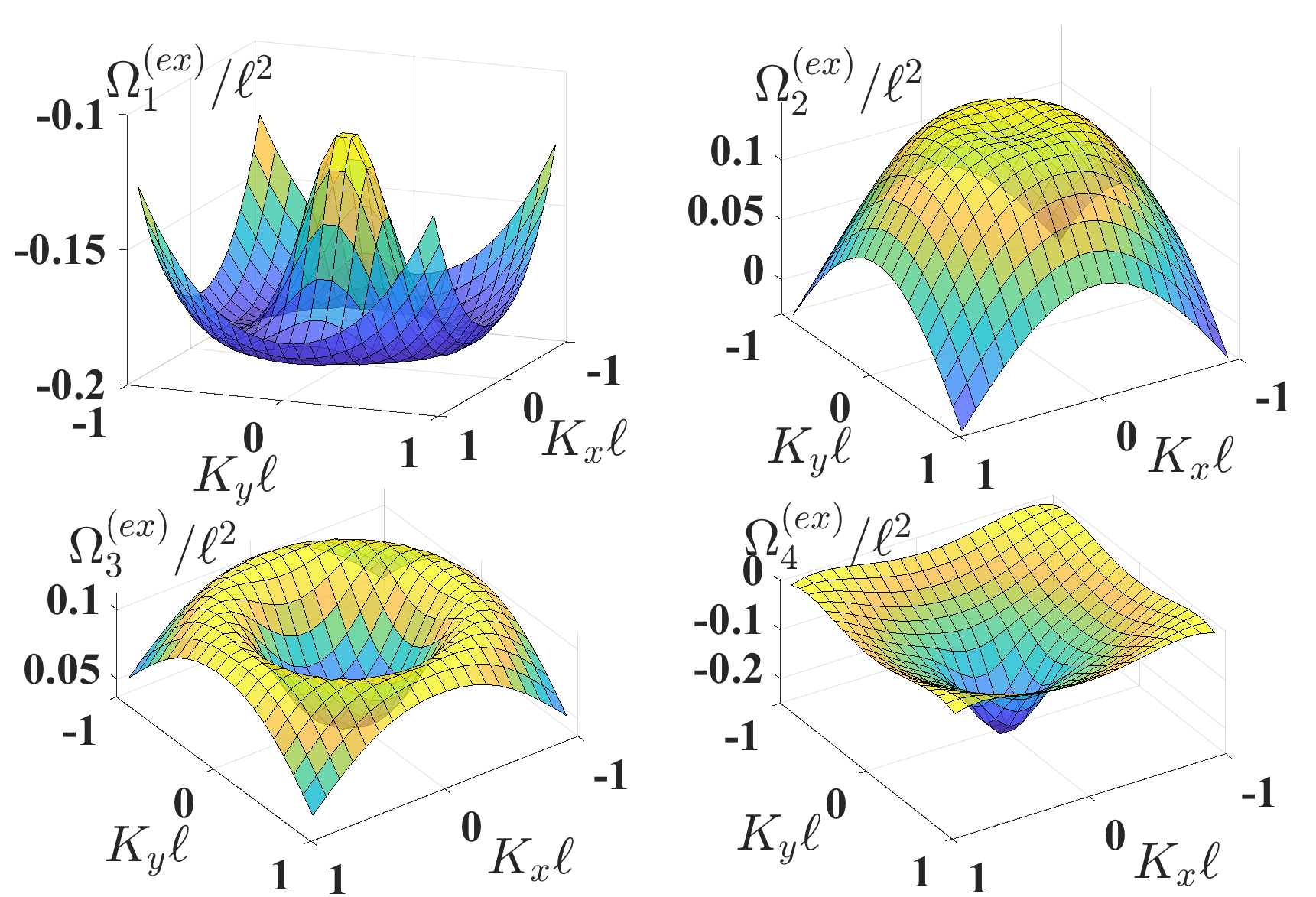}\label{fig:gappedG_BC_K_valley}}
  \hfill
  \subfloat[Exciton Berry's curvature for hole in $K'$ valley.]
  {\includegraphics[width=0.5\textwidth]{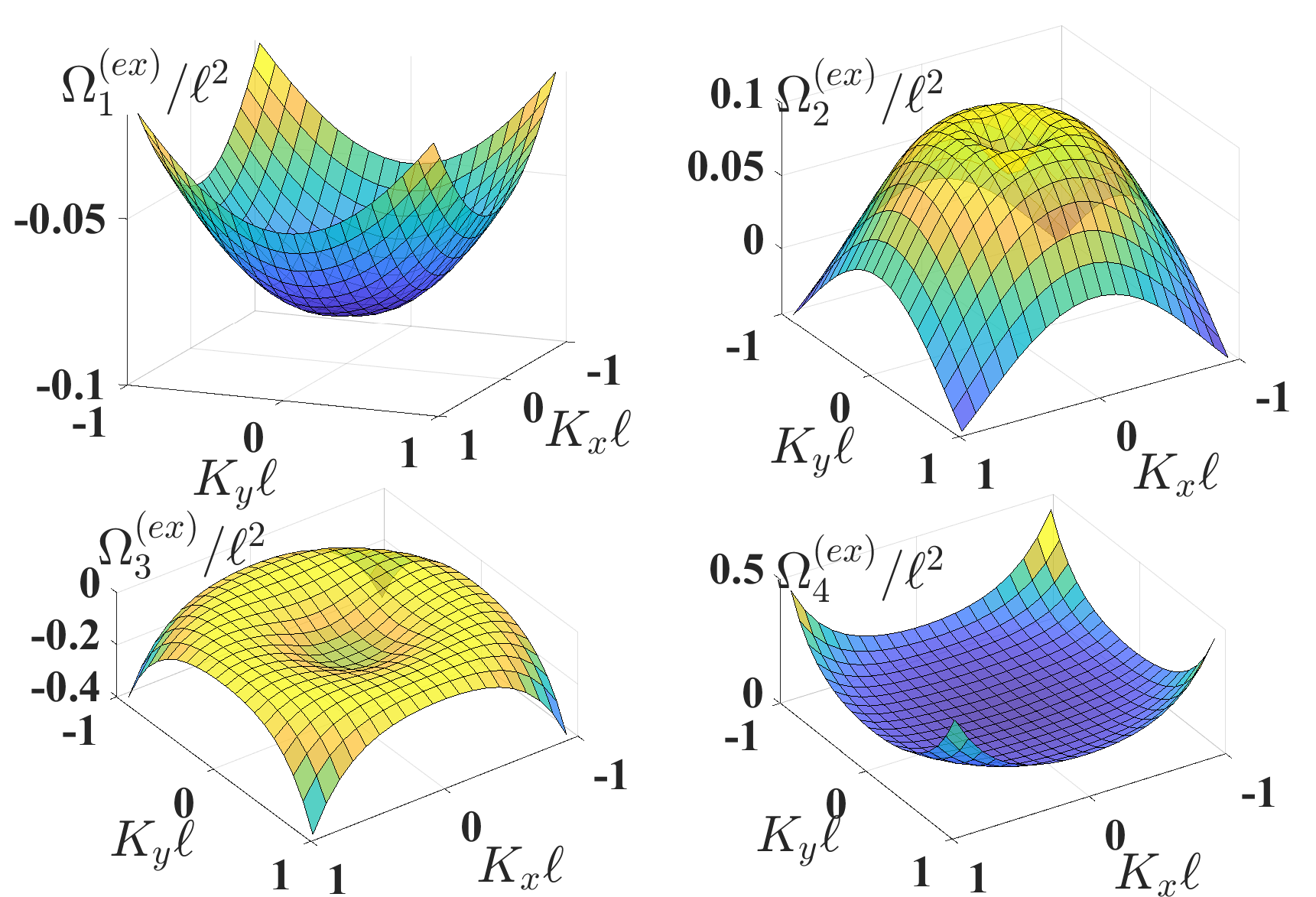}\label{fig:gappedG_BC_K'_valley}}
  \caption{Exciton Berry's curvature for lowest four exciton levels in a graphene/gapped graphene heterostructure, with hole in one or the other valley of the gapped graphene.  Parameters are those of Fig. \ref{fig:gappedG_dispersion}.
   \label{fig:gappedG_BC}
   }
\end{figure}
Fig. \ref{fig:gappedG_BC} shows the exciton Berry's curvature. For the $K$ valley [panel ({a})], the curvature for the exciton ground state ($\Omega_1^{(ex)}$) has a ``Mexican hat'' shaped surface. By contrast, the ground state exciton Berry's curvature for the $K'$ valley [panel ({b})] has a simpler bowl shape. The rich variety of behaviors apparent in the Berry's curvatures for different exciton modes, and their differences across valley, emerge from the evolution of the Landau level wavefunctions with energy.  It is interesting to notice that these behaviors are more involved than is typically found in single-particle Berry's curvatures associated with a zero-field band structure.

\subsection{TMD/TMD Heterostructure}

\begin{figure}[!tbp]
  \centering
\includegraphics[width=0.4\textwidth]{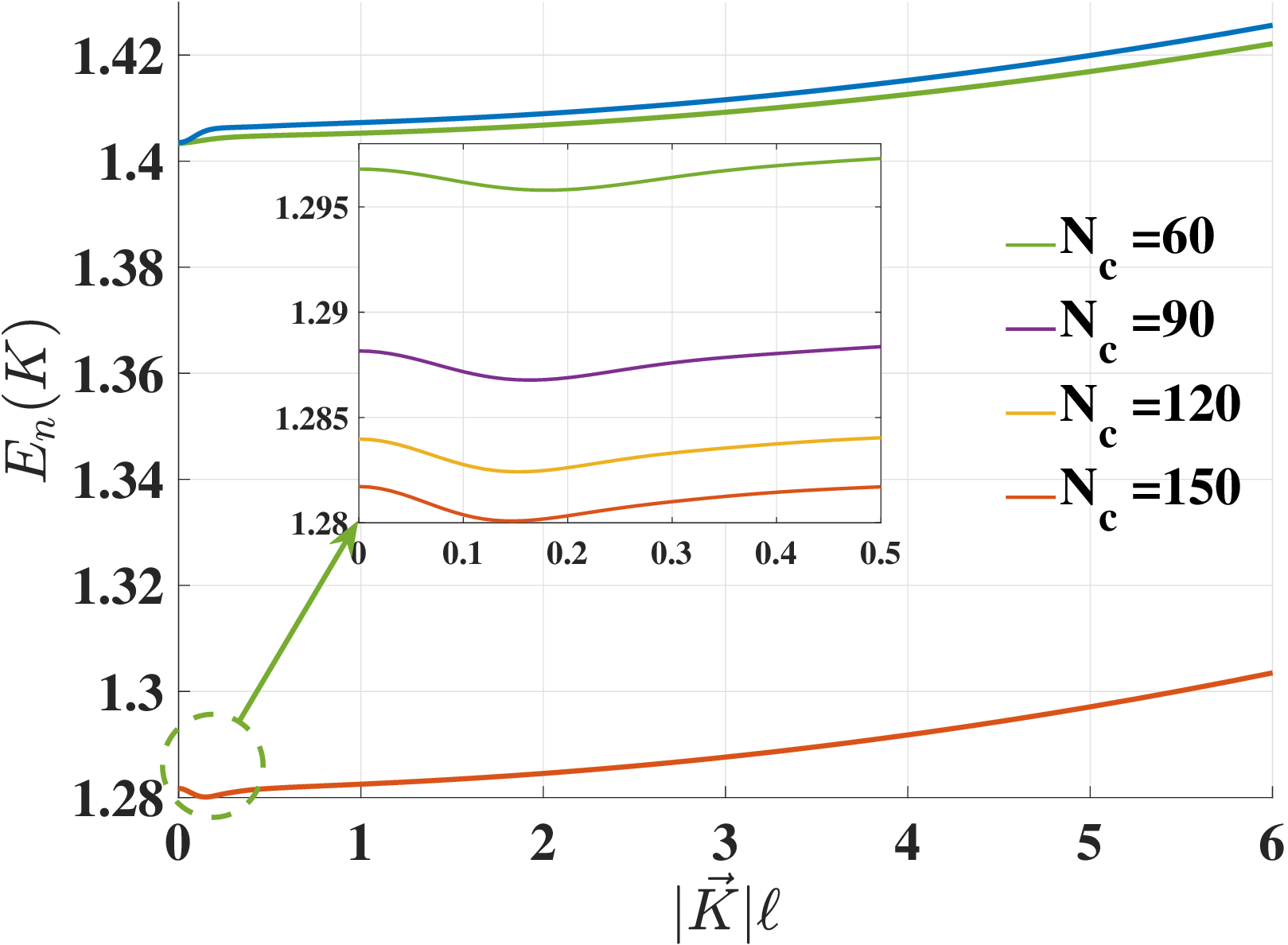}
  \caption{MoS$_2$/MoSe$_2$ heterostructure exciton dispersion in units of eV.  Parameters for non-interacting Hamiltonian are those of Ref. \onlinecite {Xiao_2012}. Hole in $K'$ valley of MoS$_2$, electron in $K'$ valley of MoSe$_2$.   Layer separation assumed at 0.71 nm, and Coulomb dielectric screening constant is 4.  Main panel illustrates results when Landau levels of index $|n| \le N_c$ are retained for both carriers, with $N_c=50$.  Inset: Evolution of lowest level energy dispersion at small momentum with varying $N_c$.}
  \label{MoS2_MoSe2_dispersion}
\end{figure}

As a final example, we consider magnetoexcitons in a heterostructure involving two different TMD materials, using  parameters corresponding to MoS$_2$ and MoSe$_2$ \cite{Xiao_2012}.
In this case we find qualitatively similar results for all combinations of valley indices of the hole and electron, and show explicit results for just one of these.
There are several features in these results which are in contrast to what we have found in our previous examples.  Figure \ref{MoS2_MoSe2_dispersion} illustrates the dispersions for the lowest three levels at low momentum.  Here a surprising result is the presence of a narrow, shallow minimum in the lowest exciton level, requiring many Landau levels to resolve.  Although our results in this case are not perfectly converged with respect to the number of Landau levels retained, we find that the position of the minimum remains relatively constant once it forms, and that it {\it deepens} as we increase our cutoff in Landau level index.  By extrapolation, we estimate that the depth of the minimum relative to the energy at zero momentum to be $\approx$ 1.8meV.  As in the case of the graphene/gapped graphene structure above, this suggests a sufficiently large density of such excitons at low temperature would condense into a state of finite momentum.  However, prior to such condensation the excitons must fall thermally into the minimum, requiring a temperature corresponding to the energy difference of this scale, $\sim$ 20 K.  Whether this can be accomplished under conditions where a large number of excitons is being generated depends on specific experimental conditions, and is a question for further study.

Beyond the behavior of the lowest level, it is notable that the first two excited exciton levels are nearly degenerate.  This is an interesting feature of this particular combination of materials.  Because of the gaps in the spectra, the Landau levels in each of the conduction and valence bands with significant weight in the wavefunctions are nearly equally spaced.  Moreover, because the Fermi velocities associated with the materials are very similar, that spacing is the same across materials.  At the non-interacting level, this results in a single lowest energy particle-hole pair, two (nearly) degenerate levels for the next level, three (nearly) degenerate levels for the third, etc.  With interactions added, these degeneracies are slightly lifted.  At higher energies, Landau levels deviate from equal spacing, and we expect to lose this clustering of the exciton modes.

\begin{figure}[!tbp]
  \centering
\includegraphics[width=0.5\textwidth]{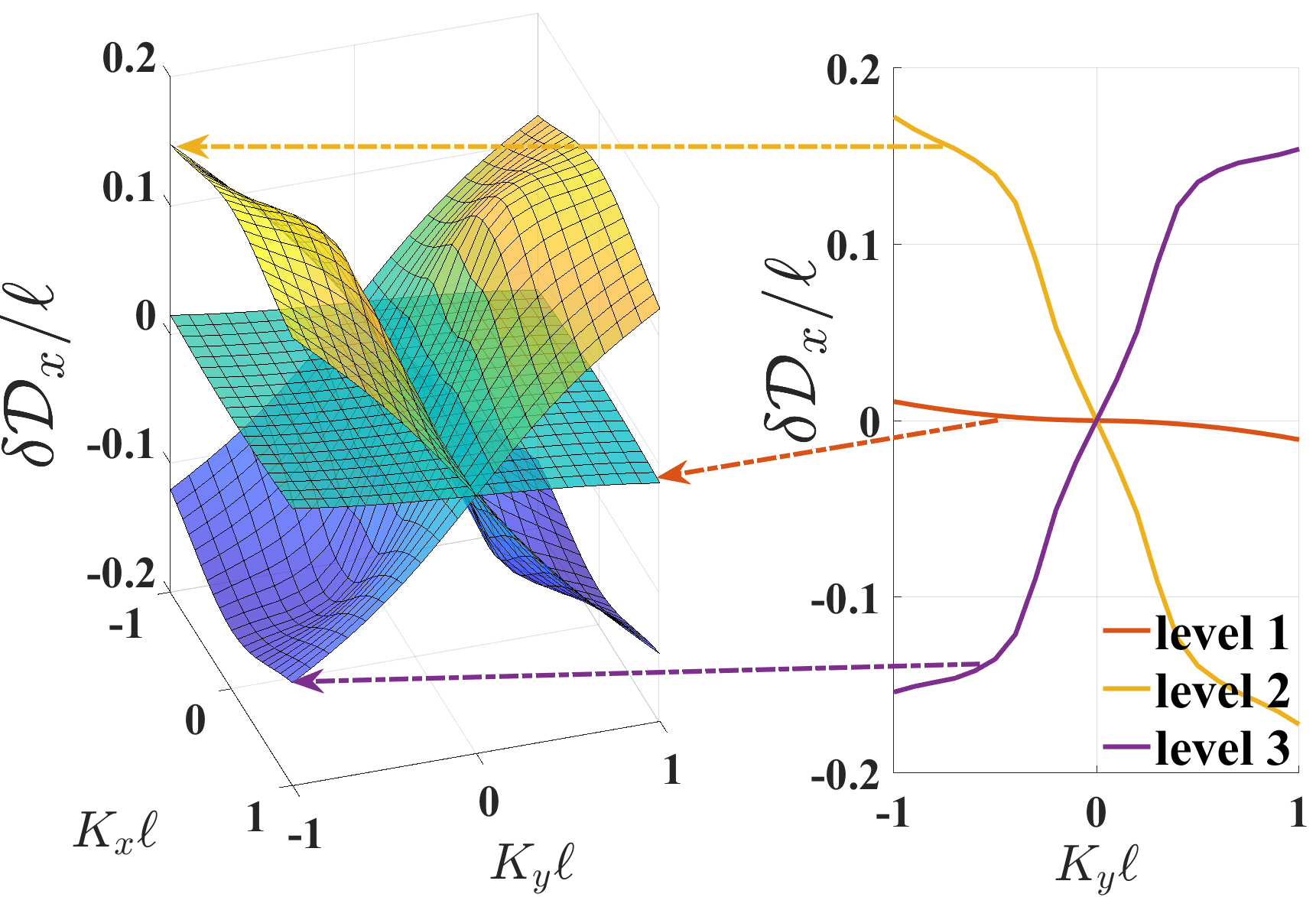}
  \caption{Correction to exciton quantum geometric dipole beyond single Landau level contribution for MoS$_2$/MoSe$_2$ heterostructure. Parameters are same as in Fig. \ref{MoS2_MoSe2_dispersion}.}
  \label{MoS2_MoSe2_dipole}
\end{figure}

Fig. \ref{MoS2_MoSe2_dipole} illustrates corrections to the exciton quantum geometric dipoles for the lowest energy levels.  Interestingly the corrections near $K=0$ are quite small for the lowest energy exciton, so that the deviation from the simple result for strong fields is relatively difficult to detect.  However, the quantum geometric effects introduced through the electron and hole band structures are much more evident in the higher energy exciton states.  Note also that the {\it sign} of the corrections varies in going from the second to third highest modes, in spite of their near degeneracy in energy.

\begin{figure}[!tpb]
  \centering
\includegraphics[width=0.7\textwidth]{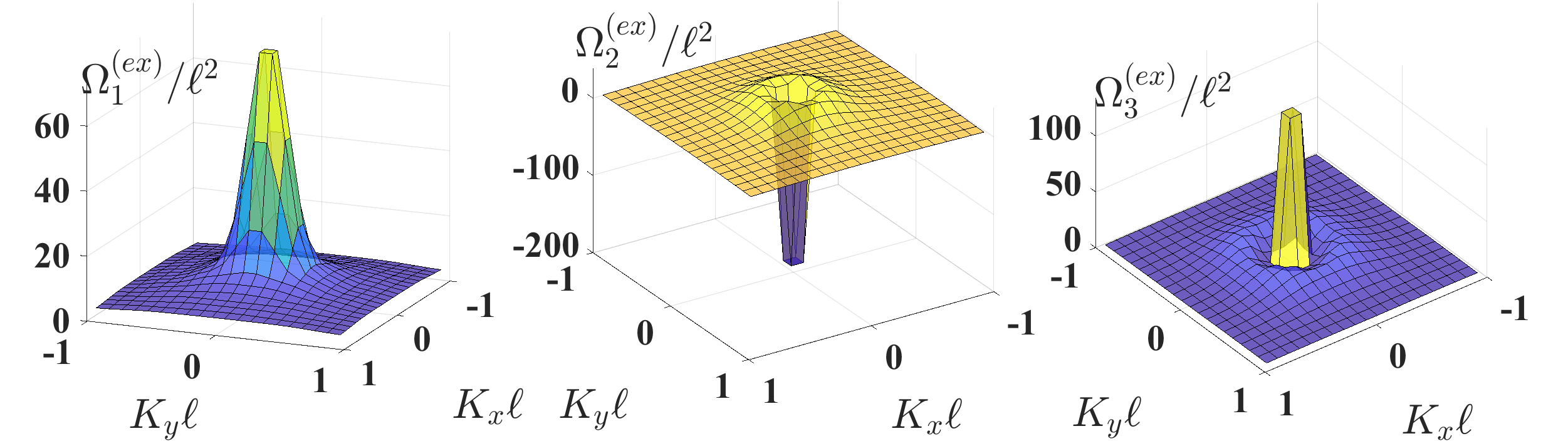}
  \caption{Exciton Berry's curvature for MoS$_2$/MoSe$_2$ heterostructure. Parameters are same as in Fig. \ref{MoS2_MoSe2_dispersion}.
  }
  \label{MoS2_MoSe2_Berry}
\end{figure}

Finally, Fig. \ref{MoS2_MoSe2_Berry} illustrates the Berry's curvatures for the lowest exciton modes.  Interestingly these have very pronounced peaks near zero momentum, significantly larger in magnitude than our corresponding results for the other heterostructures discussed above.   Together with the quantum geometric dipole results, this suggests that the semiclassical dynamics in an electric field will show band structure quantum geometric effects that are dominated by Berry's curvature \cite{Yao_2008}. Note also that the {\it signs} of these curvatures are different for different modes near zero momentum.  The striking differences in results for the different modes suggests the interesting possibility that excitons in lower and higher energy modes might be separated out spatially using electric fields, due to the significant differences in their quantum geometric dipole and Berry's curvature.  We consider this possibility in our discussions below.

\section{Discussion and Summary}
\label{section:summary_and_discussion}

\subsection{Exciton Dynamics in Electric Fields}

As a final discussion,
we turn to a qualitative analysis of the semiclassical exciton dynamics for some of the systems discussed above in static, uniform electric fields. These are governed by Eqs. \ref{Rdot_E} and \ref{Kdot_E}.  Preliminary to this, we briefly consider how excitons might be formed, and where one might look for signatures of the exciton dynamics based upon this.

Light impinging normally upon the sample plane excites intralayer excitons, which in many TMD heterostructures quickly relax into lower energy interlayer excitons \cite{Hong_2014,Yu_2015,Yu_2015b}. In such situations a collection of interlayer excitons created in a spot of initial size determined by a laser will distort with time due to imposed, in-plane electric fields.  In general excitons flow away from the central spot, distorting it in a way that can be detected in photoluminescence (PL).  By examining the spatial structure of the PL spectrum, in particular its electric field dependence, one may learn about the trajectories that the interlayer excitons have followed.  In principle, spatial- and/or time-resolved PL spectra measurements \cite{Rivera_2016,Nagler_2017,Jauregui_2019,Gisbert_2020} reflect the dynamics of different exciton modes.

If the interlayer tunneling is too slow, then intralayer particle-hole pairs may recombine before interlayer excitons form in significant numbers.  In such cases, an alternative is to directly create electron-hole pairs using light with electric field polarized normal to the plane.  We demonstrate in Appendix \ref{appendix:exciton_generation} that, provided there is some non-vanishing tunneling between the layers, this perturbation generates interlayer excitons. In materials where this creates few intralayer excitons \cite{Wang_Marie_2017}, this geometry has the advantage of eliminating a potentially large source of extraneous luminescence.  An interesting geometry in this context is to consider excitation by a narrowly focused laser {\it in} the sample plane, for which excitons would be generated along a line.  Exciton dynamics due to electric fields could then be probed by looking for PL perpendicularly away from this line.

A further possible geometry involves using van der Waals materials that do couple strongly when directly atop one another, but with a mesoscopic strip of hBN separating them so that a region of the two layers is essentially uncoupled.  In this case one may excite excitons at one edge of the hBN region with a laser, and look for signals of excitons drawn across the hBN region by an electric field on the other edge.

For the remainder of this section we assume that interlayer excitons can be created at a fixed location in a structure and consider how electric fields impact their dynamics.

\begin{figure}
  \centering
\includegraphics[width=0.8\textwidth,trim= 0 160 0 160,clip]{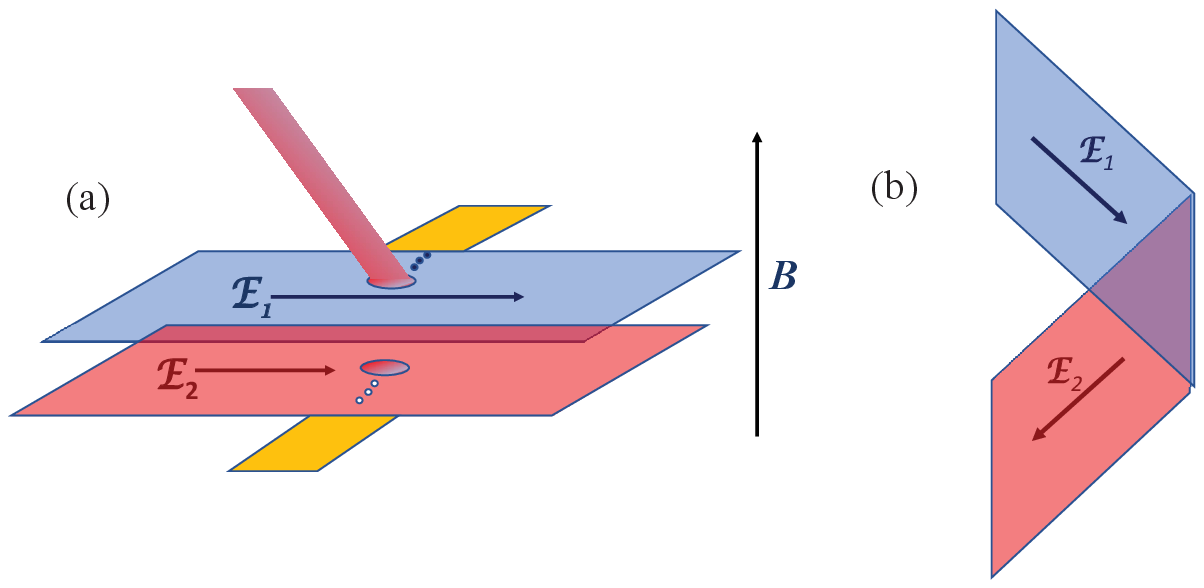}
  \caption{(a) Heterostructure with collinear applied electric fields in each layer.  A laser spot creates excitons at a location in the structure. Unequal electric fields in the two layers can cause exciton dipole moments to grow without bound, unbinding holes and electrons along trajectories perpendicular to the fields.  This could be detected as a photocurrent by contacts with separation orthogonal to the electric fields. (b)  Heterostructure with non-collinear applied electric fields.}
  \label{geometries}
\end{figure}

Consider first the simplest situation,  in which the electric fields in the two layers are collinear  [Fig. \ref{geometries}(a).]  When the fields in both layers are the same, $\dot{\bf K}_c=0$ and the exciton dynamics are determined by Eq. \ref{Rdot_E} with no contribution from the Berry's curvature.  Moreover, if the exciton is in its energetic ground state, $\vec{\nabla}_{K_c} E_0({\bf K}_c) = 0$ and the spatial dynamics is determined solely by the quantum geometric dipole.  The exciton will drift perpendicular to the electric and magnetic fields.  Deviations of this drift motion from what is expected from ${\bf v}_D ={\mathbfcal  E} \times {\bf B}/B^2$  drift in crossed fields yields a direct measure of the quantum geometric dipole.  (Note that further drift may also occur due to density or temperature gradients. However this motion should be insensitive to the electric field \cite{Onga_2017}.)In an ideal case, where this drift is ballistic, the exciton center moves at a speed $|\dot{R}_+ /2| \sim e ({\mathcal E}_+/2) |\vec{\nabla}_{K_c} {\mathcal D}({\bf K}_c)|$, where ${\mathcal E}_+/2$ is the common electric field in the two layers and ${\mathcal D}$ is the quantum geometric dipole.  For the example of the graphene/MoS$_2$ heterostructure, for the ground state exciton, with magnetic field at 10 T and an electric field of 1 V/cm, the resulting speed is $\approx 1.7v_D \sim 17 m/s.$  Using a the intervalley scattering time for a TMD heterostructure \cite{Rivera_2016} as a rough estimate of a scattering time (40 ns), this means the exciton could travel $\sim 7\mu$m before scattering, well within currently available spatial resolutions ($\sim .1 \mu$m \cite{Yuan_2020}).  If one creates excitons in higher energy modes, for the graphene-MoS$_2$ heterostructure the speed and resulting distance are reduced (see Fig. \ref{G_MoS2_K_results}), again by an amount that is well within experimental resolution.

Suppose instead one could stabilize a situation with constant ${\mathbfcal E}_-$.  Eq. \ref{Kdot_E} requires that the exciton momentum $K_c$ eventually grows without bound.  Because of the magnetic field, the centers of the electron and hole in this situation become highly separated (large dipole moment)  and the electron and hole unbind.  Since ${\mathbfcal D}$ is generally perpendicular to ${\bf K}_c$, the dissociation motion is along a direction perpendicular to the electric fields.   This situation is particularly interesting for the generation of photocurrents: in the absence of a magnetic field, the electric field must be large enough that the exciton can ionize.  In the present case they separate even at small electric fields provided the exciton dynamics remains ballistic.  Scattering by impurities or phonons, or a non-negligible moir\'e potential, impacts this result, so that in practice the time required to separate carriers must be smaller than some scattering time, setting a lower limit on the value of ${\mathbfcal E}_-$ that can ballistically separate carriers.  Finally, in addition to separation of the exciton constituents, the tendency for magnetoexcitons to have constant Berry's curvature at large $K$ (cf. Fig. \ref{G_MoS2_K_results}) means the center coordinate of the exciton will develop an anomalous constant component perpendicular to ${\mathbfcal E}_-$, even as the component parallel to it accelerates due to the $\vec{\nabla}_{K_c} E_0({\bf K}_c)$ term.  In the geometry illustrated in Fig. \ref{geometries}(a), the velocity contributions to $\dot{\bf R}_+$ from both the Berry's curvature and the quantum geometric dipole are collinear.  The two contributions can instead be orthogonal if ${\mathbfcal E_+} \perp{\mathbfcal E_-}$, allowing the two contributions to be separated.  In principle this could be achieved with a geometry such as illustrated in Fig. \ref{geometries}(b).

Clearly there are many considerations beyond the simple estimates above needed to predict the precise form of a PL spectrum for even a simple geometry.  While such modeling is beyond the scope of this work, it is interesting to note that among these are the effects of inter-exciton repulsion.  This could be modeled as an antisymmetric field ${\bf E}_-$ that would point radially outward from an initial excitation spot.  From Eq. \ref{Kdot_E} this means $\dot{\bf K}_c$ will also point radially outward, so that $\dot{\bf R}_+$ acquires a spiraling component as the initial collection of excitons begins to spread.  The presence of the dipole term in Eq. \ref{Rdot_E} allows excitons to travel away from the initial excitation spot in a partially controllable way (through the choice of ${\mathbfcal E}_+$), improving the prospects of observing the effects of dynamics in PL.

\subsection{Summary and Outlook}

In this work we have introduced a new quantum geometric quantity, the quantum geometric dipole.  The concept applies specifically to two-body systems, and exploits different choices in how a plane wave component of a wavefunction can be removed, to yield a continuum of possibilities for comparing wavefunctions at different wavevectors ${\bf K}$ via a scalar product.  Differences among these scalar products are gauge-invariant quantities, and we demonstrated that one particular choice is directly proportional to the average displacement vector between the two constituents of the wavefunction.  In the case of two oppositely charged particles, this yields the dipole moment as a function of ${\bf K}$.

Excitons in two-dimensional semiconductors represent one class of relevant systems.  We largely focused our attention on double layer heterostuctures in magnetic fields, with hole and electrons residing in different layers.   The presence of a magnetic field opens Landau level gaps in the layer spectra so that systems with graphene constituents can be considered.  Such heterostructures are particularly interesting because the holes and electrons may reside in bands with very different Berry's curvatures, allowing one to examine how these impact results for the excitons as a whole.

An examination of exciton wavepacket dynamics demonstrated that, in addition to the exciton dispersion as a function of momentum, both Berry's curvature and the quantum geometric dipole enter into the equations of motion. The latter two couple to uniform electric fields in the layers in simple ways.  We then examined the relevant quantities for a few representative systems: gapped graphene (in zero field), graphene on a TMD surface (along with a simpler gapped graphene model of the latter), and TMD surfaces of different materials.  In some cases we found that the energy dispersions had their lowest energies at a non-vanishing $|{\bf K}|$, even for the lowest energy exciton states.  In a strong magnetic field, the quantum geometric dipole implies an exciton dipole moment with a well-known form, but the effects of band structure for weaker fields results in considerable corrections to this.  Our calculations for systems in magnetic fields show that the quantum geometric dipole increases in magnitude to large values, while the Berry's curvature tends to vanish, at large momentum.  In our zero field example these both vanish at large momentum.  By examining results for a gapped/gapless graphene heterostructure, we found that differing orbital structures of wavefunctions with energy will lead to significant differences in relevant quantities with respect to which valleys the constituents reside in.  Orbital structures that remain relatively constant are largely insensitive to the choices of valley, although an overall shift in the exciton dispersions can be sizeable.

Our analysis above suggests further interesting directions for study.  Among these are understanding the effects of the full moir\'e band structure on exciton behavior, the effects of inter-exciton interactions, many-body behavior beyond the simple two-body approximation adopted in this study, and more detailed modeling of exciton distributions and their responses to electric fields.  Beyond these, a further direction becomes relevant
for a sufficiently high density of optically-generate excitons: at low temperature there is the possibility that they will Bose-condense \cite{Wang:2019aa}.  Indeed,
finite densities of excitons in these systems could be induced by an interlayer bias:  a perpendicular electric field impacts the single particle-energies of the hole and the electron separately, shifting the entire excitation curve up or down.  If part of the exciton curve becomes negative,  one expects charge transfer between the layers to create a finite density of excitons, again opening the possibility of Bose condensation \cite{Paquet_1985,Fertig_1989,Zhu_2019}.  When the lowest exciton dispersion has its minimum at finite momentum, such condensation occurs into a state with finite electric dipole moment.  Coherent dynamics of these condensed excitons offers unique possibilities to observe quantum geometric properties of their band structure \cite{Yao_2008}.

The presence of an in-plane dipole moment associated with magnetoexcitons in these systems is highly reminiscent of an $XY$ model with dipolar interactions, the ``dipolar $XY$'' model.  The relatively long-range interactions of dipolar interactons makes this system distinct from the more standard $XY$ model.  It is believed to have an ordered ``ferromagnetic'' groundstate (really ferroelectric for excitons), and a Berezinskii-Kosterlitz-Thouless transition to a dipole disordered state \cite{Maier_2004,Vasiliev_2014}.  The Bose-condensed state thus has two-broken continuous symmetries, one associated with the dipole direction and  another with the condensate phase.  In the exciton system, the dynamics of these order parameters are coupled in a non-trivial way because of the relation between electric dipole moment and the momentum into which the excitons condense, suggesting that thermal disordering of this system may be quite unique.

Quantum geometric phases are at the root of this coupling.  Studies of such excitons, particularly as realized in certain heterostructure systems, offer unique windows into the effects of these phases on the quantum physics of many-body systems.


\vskip 0.5cm

\textit{Acknowledgements --}
LB acknowledges support by Spain's MINECO under Grants No. PGC2018-097018-B-100.
HAF and JC acknowledge the support of the NSF  through Grant Nos. DMR-1914451 and ECCS-1936406.
HAF acknowledges support from the US-Israel
Binational Science Foundation (Grant Nos. 2016130 and 2018726),
of the Research Corporation for Science Advancement through a Cottrell SEED Award, and
the hospitality and support of the Aspen Center for Physics (Grant No.
PHY-1607611), where part of this work was done.  Computational work was supported in part by the Lilly Endowment, Inc., through its support for the Indiana University Pervasive Technology Institute.

\newpage
\appendix

\section{Derivation of Effective Lagrangian}
\label{appendix:lagrangian}

As discussed in the main text, the general form for the Lagrangian is
\begin{equation}
\label{lagrangian_appendix}
{\mathcal L} = \langle \Psi |i\partial_t | \Psi\rangle - \langle \Psi | H | \Psi\rangle \equiv {\mathcal L}_t - E({\bf K}),
\end{equation}
where the state $|\Psi\rangle$ in the absence of external fields is given by $|\Psi_0\rangle$, Eq. \ref{wavepacket}.
Due to the additional fields,
Eq. \ref{wavepacket} is modified to \cite{Chang_1996}
\begin{equation}
\label{Psidef}
\langle {\bf r}_1,{\bf r}_2 | \Psi\rangle \approx e^{-ie\delta A^{(2)}({\bf R}_2,t) \cdot
{\bf r}_2 + ie\delta {\bf A}^{(1)}({\bf R}_1,t) \cdot {\bf r}_1} \langle {\bf r}_1,{\bf r}_2 | \Psi_0 \rangle.
\end{equation}
It is assumed that the fields encoded in $\delta {\bf A}^{(i)}$ vary slowly over the real space size of the wavepacket.
Assuming for simplicity that the electric and addition magnetic field are spatially uniform, we can write $\delta {\bf A}^{(i)} ({\bf r},t) =  - {\mathbfcal E}^{(i)} t + {1 \over 2} \delta {\mathbfcal B} \times {\bf r}$.  With some algebra one finds
\begin{equation}
\label{lagrangian_E}
E({\bf K}) = E_0({\bf K}) - {e \over 2} \delta {\mathbfcal B} \cdot
\langle \Psi | \left( {\bf r}_1 -{\bf R}_1 \right) \cdot \vec{V}^{(1)} - \left( {\bf r}_2 -{\bf R}_2 \right) \cdot \vec{V}^{(2)} | \Psi \rangle,
\end{equation}
where $E_0({\bf K})$ is the energy of the exciton at momentum ${\bf K}$ (in the state $|\Phi_{\bf K}\rangle$) in the absence the electric and additional magnetic fields.

Note that in this formulation, the instantaneous energy $E({\bf K})$ has no dependence on the electric fields; these instead appear in ${\mathcal L}_t$, as we now show. Using Eqs. \ref{wavepacket} and \ref{Psidef}, and the definition in Eq. \ref{lagrangian}, one may write
\begin{equation}
{\mathcal L}_t = e\langle \Psi_0 | \delta \dot{\bf A}^{(2)}({\bf R}_2,t) \cdot {\bf r}_2 - \delta \dot{\bf A}^{(1)}({\bf R}_1,t) \cdot {\bf r}_1 |\Psi_0 \rangle + i\int d^2K w^*({\bf K}) \partial_t w({\bf K}),
\end{equation}
where $\dot{\bf A}$ indicates a time derivative of ${\bf A}$.
Writing $w({\bf K})=|w({\bf K})|e^{-i\gamma({\bf K},t)}$, and assuming the time-dependent part of the phase $\gamma$ varies slowly in space, one finds
\begin{equation}
\label{lt1}
{\mathcal L}_t \approx e\delta \dot{\bf A}^{(1)} \cdot {\bf R}_1-e\delta \dot{\bf A}^{(2)} \cdot {\bf R}_2 + \partial_t \gamma({\bf K}_c,t).
\end{equation}
Our goal is to treat ${\bf R}_1$, ${\bf R}_2$, and ${\bf K}_c$ as collective degrees of freedom and write down equations of motion for them.  Formally this is accomplished by writing \cite{Marder_book} an action $S=\int dt {\mathcal L}$ and minimizing with respect to the these parameters.  Thus, up to total derivative terms which will not impact the equations of motion, one may rewrite Eq. \ref{lt1} as
\begin{equation}
\label{lt2}
{\mathcal L}_t \approx -e\delta {\bf A}^{(1)} \cdot \dot{\bf R}_1+e\delta {\bf A}^{(2)} \cdot \dot{\bf R}_2 - \dot{\bf K}_c \cdot \vec{\nabla}_{K_c} \gamma({\bf K}_c).
\end{equation}
As discussed in the main text,  Eqs. \ref{position_connection_wp_hole}, \ref{position_connection_wp_electron} imply a constrain, Eq. \ref{constraint},
so that ${\bf R}_1$ and ${\bf R}_2$ are not independent.  Using these equations,
and adopting ${\bf R}_+={\bf R}_1+{\bf R_2}$ and ${\bf K}_c$ as our independent degrees of freedom,
the Lagrangian may be cast in the form
\begin{align}
{\mathcal L} &= {\mathcal L}_t - E({\bf K}_c) \nonumber \\
&= {e \over 2} \sum_{\mu=x,y} \delta A_{+,\mu} \dot{\bf K}_c \cdot \vec{\nabla}_{K_c} \left[{\mathcal A}^{(1)}_{\mu}-{\mathcal A}^{(2)}_{\mu} \right] - {e \over 2} \delta {\bf A}_- \cdot \dot{\bf R}_+ -{1 \over 2} \dot{\bf K}_c \cdot \left[{\bf R}_+
-{\mathbfcal A}^{(1)}-{\mathbfcal A}^{(2)} \right]-E_0({\bf K_c}),
\label{lagrangian_final}
\end{align}
where we have defined $\delta A_{\pm} = \delta A^{(2)} \pm \delta A^{(1)},$ as presented in Eq. \ref{lagrangian_main_text} in the main text.

\newpage

\section{Eigenvalue Equation for Excitons}
\label{appendix:exciton_equations}

In this Appendix we provide a more detailed presentation of the eigenvalue equations we solve to compute the properties of interlayer magnetoexcitons.

\subsection{Single and Two-Particle Wavefunctions}

Our single-particle Hamiltonians are assumed to be generically of the form in Eq.
\ref{eq: single layer Hamiltonian general form}, or simple variants of this that incorporate the effects of spin, which is of particular interest in the TMD case \cite{Xiao_2012}.  To proceed we will require energies and wavefunctions of these Hamiltonians in the presence of a magnetic field.  For both types of systems, these have been worked out previously; for completeness we review the relevant results.


To introduce an orbital magnetic field into the Hamiltonian of Eq. \ref{eq: single layer Hamiltonian general form} for an electron, one makes the Peierls substitution ${\bf p} \rightarrow \mathbf{\Pi}={\bf p}+e{\bf A}$, where ${\bf A}$ is the vector potential for a uniform magnetic field.  Note that for $e>0$, the vector potential has been added as appropriate for a negatively charged particle.
%
%
The simplest case is that of graphene \cite{Gusynin_2005,Goerbig_2011}.  Its
positive energy states have the form
\begin{equation}
\vec{\xi}_{nk\tau=+1}^{(e)}=
\left(
\begin{array}{c}
\alpha^{(g)}_{n,\sigma,\tau=1} \psi_{n-1,k}^{(-1)} \\
\beta^{(g)}_{n,\sigma,\tau=1}\psi_{n,k}^{(-1)}\end{array}
\right),
\quad
\quad
\vec{\xi}_{nk\tau=-1}^{(e)}=
\left(
\begin{array}{c}
\alpha^{(g)}_{n,\sigma,\tau=-1}\psi_{n,k}^{(-1)} \\
\beta^{(g)}_{n,\sigma,\tau=1}\psi_{n-1,k}^{(-1)}\end{array}
\right),
\label{graphene_wf}
\end{equation}
where $\alpha^{(g)}_{n,\sigma,\tau}=\tau/\sqrt{2}$, $\beta^{(g)}_{n,\sigma,\tau}=1/\sqrt{2}$ for $n \ge 1$; for $n=0$, $\alpha^{(g)}_{n=0,\sigma,\tau=1}=0$,  $\beta^{(g)}_{n=0,\sigma,\tau=1}=1$, $\alpha^{(g)}_{n=0,\sigma,\tau=-1}=1$, and $\beta^{(g)}_{n=0,\sigma,\tau=1}=0$.
The wavefunctions $\psi^{(s=-1)}_{nk}$ are given by Eq. \ref{LLstates}, and the energies of these states are $\varepsilon_n^{(e)}=\sqrt{2n}v_G/\ell$.

The Landau level wavefunctions for TMD materials near the $K$ and $K'$ points are somewhat more involved, but also may be written explicitly \cite{Li_2013}.
We first consider single-particle wavefunctions for an electron in the conduction band (positive energy states), for which
\begin{equation}
  \vec{\psi}^{(e)}_{n,k,\sigma,\tau=+1} =
  \left(
    \begin{array}{c}
      \alpha^{(e)}_{n,\sigma,\tau=1}\psi^{(-1)}_{n-1,k} \\
      \beta^{(e)}_{n,\sigma,\tau=1} \psi^{(-1)}_{n,k} \\
    \end{array}
  \right),
  \quad
  \vec{\psi}^{(e)}_{n,k,\sigma,\tau=-1} =
  \left(
    \begin{array}{c}
      \alpha^{(e)}_{n,\sigma,\tau=-1}\psi^{(-1)}_{n,k} \\
      \beta^{(e)}_{n,\sigma,\tau=-1}\psi^{(-1)}_{n-1,k} \\
    \end{array}
  \right),
  \label{TMD_wf_e}
\end{equation}
where $\sigma$ denotes the spin state of the electron.
In these expressions, $n \ge 1$.  In addition, there is an $n=0$ Landau level for positive energy states in the $K'$ $(\tau=-1)$ valley.  In these expressions $\sigma=\pm 1$ indicates the spin state of the electron.  {The coefficients $\alpha^{(e)}$, $\beta^{(e)}$, as well as explicit expressions for their energies $\varepsilon^{(e,TMD)}_{n,\sigma,\tau}$, are provided at the end of this Appendix.} Depending on the material hosting the electron, the wavefunctions $\xi^{(e)}$ or $\psi^{(e)}$ play the role of $\psi^{(s=-1)}$ in Eq. \ref{ewfsc} to form states with good two-dimensional momentum quantum numbers.

In addition to these wavefunctions for the electron, we require a basis of states for the hole.  In the examples we present, we assume the hole lies among the negative energy states of a TMD material.  In the two-body approximation we have adopted, the hole is modeled as positively charged particle residing in a linear combination of the particle-hole conjugates of these states.  Thus, their wavefunctions are composed from states of the form
\begin{equation}
  \vec{\psi}^{(h)}_{n,k,\sigma,\tau=+1} =
  \left(
    \begin{array}{c}
      \alpha^{(h)}_{n,\sigma,\tau=1}\psi^{(+1)}_{n,k} \\
      \beta^{(h)}_{n,\sigma,\tau=1} \psi^{(+1)}_{n-1,k} \\
    \end{array}
  \right),
  \quad
  \vec{\psi}^{(h)}_{n,k,\sigma,\tau=-1} =
  \left(
    \begin{array}{c}
      \alpha^{(h)}_{n,\sigma,\tau=-1}\psi^{(+1)}_{n-1,k} \\
      \beta^{(h)}_{n,\sigma,\tau=-1}\psi^{(+1)}_{n,k} \\
    \end{array}
  \right),
  \label{TMD_wf_h}
\end{equation}
with corresponding energies $\varepsilon^{(h)}_{n,\sigma,\tau}$ which are taken as positive.    Again the integers $n \ge 1$, but in addition there is an $n=0$ state, this time in $\tau=+1$ valley.
We again relegate details of the {the coefficients $\alpha^{(h)}$, $\beta^{(h)}$ and the single particle energies $\varepsilon^{(h)}_{n,\sigma,\tau}$, to the end of this Appendix.}

Since we are ultimately interested in geometric phases related to changes in momentum, we wish to recast these states in forms that are eigenstates of magnetic translations, Eqs. \ref{e_MTO}.  This is accomplished by taking linear combinations precisely of the form in Eq. \ref{ewfsc}, yielding states with well-defined two-dimensional momenta {\bf q}, which we write in the form $\vec{\phi}^{(h)}_{n_h,{\bf q},\sigma,\tau}$ for the hole, and $\vec{\phi}^{(e)}_{n_e,{\bf q},\sigma,\tau}$ for the electron.  Note that in these studies the hole states involve linear combinations of $\vec{\psi}^{(h)}_{n,k,\sigma,\tau}$, while the for the electron we will consider wavefunctions involving either $\vec{\xi}^{(e)}$ or $\vec{\psi}^{(e)}$, depending on whether it resides in graphene or a TMD material.

Our trial wavefunctions for the exciton generically are analogous to Eq. \ref{exciton_wf1},
\begin{equation}
\Phi_{\bf K} = \sum_{\bf{n}\bf{q}} C_{{\bf n}}({\bf K}) e^{i{\bf q}\cdot {\bf R}}e^{iK_x^{(0)}q_y \ell^2} \vec{\phi}^{(h)}_{n_h,{\bf q},\sigma_h,\tau_h}({\bf r}_1) \otimes \vec{\phi}^{(e)}_{n_e,{\bf K^{(0)}}-{\bf q},\sigma_e,\tau_e}({\bf r}_2),
\end{equation}
where the full momentum ${\bf K}={\bf K}^{(0)} -\hat{z} \times {\bf R}/\ell^2$, has been broken up into a part that lies in the first Brillouin zone, ${\bf K}^{(0)}$, and a reciprocal lattice vector, $-\hat{z} \times {\bf R}/\ell^2$,
where ${\bf R}$ is a direct lattice vector, as discussed in Sec. \ref{section:LandauLevels}.
The vector ${\bf n} \equiv (n_1,n_2)$ represents the set of possible Landau level indices the hole ($n_1$) and the electron ($n_2$) may have (i.e., the set of levels which are not Pauli-blocked by other carriers.)

\subsection{Eigenvalue Equation and Berry's Connections}

In analogy with Eqs. \ref{H2body} and \ref{ggHam}, the Hamiltonian of the two-body system has the form $H=H^{(+)}({\bf r}_1) \otimes \mathbbm{1} + \mathbbm{1} \otimes H^{(-)}({\bf r}_2) + v({\bf r}_1 - {\bf r}_2)$, and we wish to minimize $\langle \Phi_{\bf K} | H | \Phi_{\bf K}\rangle$ with respect to the coefficients $C_{{\bf n}}({\bf K})$, subject to the constraint that $|\Phi_{\bf K}\rangle$ is normalized.  This leads to the equation
\begin{align}\label{eval_equation_generic}
 \sum_{{\bf n}'} \left\{\frac{1}{(2\pi)^2}\int d^2k' \widetilde{v}(k') S^{(h),n_1,n_1'}_{{\bf q},{\bf q}+{\bf k}'}({\bf k}') S^{(e),n_2,n_2'}_{{\bf K^{(0)}} - {\bf q},{\bf K^{(0)}}-({\bf q}+{\bf k}')}({\bf k}')  e^{i{\bf k}' \cdot {\bf R}}
 + \varepsilon^{(h)}_{n'_1}+\varepsilon^{(e)}_{n'_2} \right\} C_{{\bf n}'}
            = E({\bf K}) C_{{\bf n}},
\end{align}
where $\varepsilon_n^{(h,e)}$ are the single-particle energies for the hole and electron.  The structure factors,
\begin{equation}
S_{{\bf q},{\bf q}'}^{(h),n,n'}({\bf k}) = \langle {\phi}^{(h)}_{n,{\bf q},\sigma,\tau} | e^{-i{\bf k} \cdot {\bf r}} | {\phi}^{(h)}_{n',{\bf q}',\sigma,\tau} \rangle,
\quad
S_{{\bf q},{\bf q}'}^{(e),n,n'}({\bf k}) = \langle {\phi}^{(e)}_{n,{\bf q},\sigma,\tau} | e^{i{\bf k} \cdot {\bf r}} | {\phi}^{(e)}_{n',{\bf q}',\sigma,\tau} \rangle,
\label{struc_fac_hole}
\end{equation}
are linear combinations of those appearing in Eq. \ref{Bloch_overlap_general}, with weights that are easily read off from the various wavefunctions.  For every value of ${\bf K}$, Eq. \ref{eval_equation_generic} provides us with a discrete matrix equation to solve that yields the energies and wavefunctions of the exciton states a particle-hole pair supports at a given momentum.

Several comments are in order. First, in this formulation, the number of states at a given ${\bf K}^{(0)}$ will be equal to the product of the number of ${\bf R}$ vectors and the number of Landau level pairs (for particle and hole) retained.   The energy $E({\bf K})$ may be represented in an extended zone scheme by plotting the energy of an eigenvalue for a given ${\bf R}$ at ${\bf K}^{(0)} -\hat{z} \times {\bf R}/\ell^2$.
Because we are not including any periodic potential or tunneling in our model Hamiltonian, dispersions for neighboring ${\bf R}$'s will always connect in a continuous fashion \cite{Marder_book}. Secondly, in general the integral appearing in Eq. \ref{eval_equation_generic} must be evaluated numerically, and in practice this turns out to be the limiting step in our numerical diagonalization scheme.  Finally, the number of Landau level pairs retained impacts the accuracy of our solutions.  For most cases we are able to obtain quantitative results for the lowest few exciton dispersions with sufficiently large numbers of these.  However, for the case of TMD/TMD heterostructures, our results are semi-quantitative, although the shapes of the exciton dispersions as well as the computed curvatures are qualitatively correct.

The eigenvalues in Eq. \ref{eval_equation_generic} provides one element of what is needed in the equations of motion in Section \ref{semiclassical}.  Beyond this we need to compute the quantum geometric dipole and the Berry's curvature.  These are obtained from the wavefunctions using the Berry's connections, ${\mathbfcal A}^{(1)}$ and ${\mathbfcal A}^{(2)}$.  For the wavefunctions as we have written them, one finds after some algebra
\begin{align}
\label{A1}
{\mathbfcal A}^{(1)} &= i\sum_{\bf n} C^*_{\bf n}({\bf K}) \vec{\nabla}_{\bf K} C_{\bf n}({\bf K}) - K_x\hat{y} - \sum_{{\bf n},{\bf n}'} C^*_{\bf n}C_{{\bf n}'}\left[-D^{(h)}_{n_1,n_1'}\hat{x} - i{\rm sgn}(n_1'-n_1)D^{(h)}_{n_1,n_1'}\hat{y} \right] \delta_{n_2,n_2'},
\\
\label{A0}
{\mathbfcal A}^{(2)} &= i\sum_{\bf n} C^*_{\bf n}({\bf K}) \vec{\nabla}_{\bf K} C_{\bf n}({\bf K}) - K_y\hat{x} - \sum_{{\bf n},{\bf n}'} C^*_{\bf n}C_{{\bf n}'}\left[-D^{(e)}_{n_2,n_2'}\hat{x} + i{\rm sgn}(n_2'-n_2)D^{(e)}_{n_2,n_2'}\hat{y} \right] \delta_{n_1,n_1'},
\end{align}
with
\begin{equation}
\label{Ddef}
D_{n,n'}^{(\mu)} \equiv \left(\alpha_n^{(\mu)}\alpha_{n'}^{(\mu)}\sqrt{\frac{\max(n,n')-1}{2}} + \beta_{n}^{(\mu)}\beta_{n'}^{(\mu)}\sqrt{\frac{\max(n,n')}{2}} \right) \left(\delta_{n',n+1}+\delta_{n',n+1} \right).
\end{equation}
In this last equation, $\mu$ is either $h$ or $e$ for the electron or hole, but the $\alpha$ and $\beta$ coefficients must actually be drawn from Eqs. \ref{graphene_wf}, \ref{TMD_wf_e}, or \ref{TMD_wf_h}, depending on what material the carrier is residing in.  (For brevity we have suppressed the $\sigma$ and $\tau$ indices in these coefficients.)

As mentioned above these are gauge-dependent quantities.  Their difference yields the (gauge-independent) quantum geometric dipole, which may be evaluated straightforwardly from the numerical solutions of Eq. \ref{eval_equation_generic}.  Note that in forming this difference, the first terms of Eqs. \ref{A1} and \ref{A0} cancel, while the second terms combine to give the single Landau level quantum geometric dipole, with an associated drift velocity in an electric field that can be understood completely in terms of a simple velocity boost to a frame in which the field vanishes.   The last terms give the corrections to this coming from non-trivial structure in the wavefunctions; if the electron and hole states are each confined to a single Landau level, this contribution vanishes.

Calculation of the Berry's curvature is more involved because it involves a curl of the sum of these quantities.  Our numerical diagonalizations do not typically yield results that vary smoothly with ${\bf K}$, creating a barrier for differentiation.  We circumvent this problem in a standard way: instead of evaluating the Berry's curvature directly, we evaluate its {\it flux} through a grid of small plaquettes, and divide by the plaquette area.  Each of the fluxes are formally equal to a line integral around a plaquette boundary, and for small plaquettes these are well-approximated in terms of (the logarithm of) products of overlaps between wavefunctions on the plaquette corners \cite{Vanderbilt_book}.  The results converge to the Berry's curvature in the limit of very small plaquette size.

\subsection{Coefficients for TMD Wavefunctions}
\label{section:coefficients}

Finally, we provide the remaining details of the wavefunctions used in our calculations involving TMD materials and gapped graphene. The wavefunctions have the same form in both these cases, given by Eq. \ref{TMD_wf_e}, but with $\lambda$ set to zero for gapped graphene.  For Landau levels with $n\neq0$ and in the $K$ valley ($\tau = +1$), the $\alpha$ and $\beta$ coefficients are 
\begin{align}\label{alpha beta coefficients K valley}
  &\alpha^{(e)}_{n,\sigma,\tau=1} = \beta^{(h)}_{n,\sigma,\tau=1} = N_{n\sigma +}^{} \frac{\sqrt{2n}v_F\hbar}{\ell}, \\
  &\beta^{(e)}_{n,\sigma,\tau=1} = \alpha^{(h)}_{n,\sigma,\tau=1} = N_{n\sigma+}^{} \left(-\frac{\Delta -\sigma \lambda}{2} + \sqrt{ \left(\frac{\Delta - \sigma \lambda}{2}\right)^2 +\frac{v_F^2 \hbar^2}{\ell^2}2n} \right), \\
  &N_{n \sigma +}^{}  = \left( 4n \left(\frac{v_F\hbar}{\ell}\right)^2+\frac{(\Delta- \sigma \lambda)^2}{2} - (\Delta - \sigma \lambda)\sqrt{ \left(\frac{\Delta - \sigma \lambda}{2}\right)^2 +\frac{v_F^2 \hbar^2}{\ell^2}2n } \right)^{-\frac{1}{2}}.
\end{align}
For this valley, there are no $n=0$ states.


In the $K'$ valley, the states have the coefficients
\begin{align}\label{alpha beta coefficients K' valley}
  &\alpha^{(e)}_{n,\sigma,\tau=-1} =\beta^{(h)}_{n, \sigma,\tau=-1}= N_{n\sigma -}^{} \left(\frac{\Delta +\sigma \lambda}{2} + \sqrt{ \left(\frac{\Delta + \sigma \lambda}{2}\right)^2 +\frac{v_F^2 \hbar^2}{\ell^2}2n} \right) , \\
  &\beta^{(e)}_{n, \sigma,\tau=-1} = \alpha^{(h)}_{n,\sigma,\tau=-1} = - N_{n \sigma -}^{} \frac{\sqrt{2n}v_F\hbar}{\ell}, \\
  &N_{n \sigma -}^{}  = \left( 4n \left(\frac{v_F\hbar}{\ell}\right)^2+\frac{(\Delta+ \sigma \lambda)^2}{2} + (\Delta + \sigma \lambda)\sqrt{ \left(\frac{\Delta + \sigma \lambda}{2}\right)^2 +\frac{v_F^2 \hbar^2}{\ell^2}2n } \right)^{-\frac{1}{2}},
\end{align}
For this valley these expression apply for $n \ge 0$.

The energies of these various states are given by the single expression
\begin{equation}\label{MoS2 energy band K' valley}
  \varepsilon^{(e,h)}_{n,\sigma,\tau} = \pm \frac{\sigma\tau\lambda}{2} + \sqrt{ \left( \frac{ \Delta - \sigma \lambda \tau}{2}  \right)^2  + \frac{v_F^2\hbar^2}{\ell^2} 2n },
\end{equation}
where the upper corresponds to the electron, and the lower to the hole.
Again, for gapped graphene, this same expression applies, with $\lambda \rightarrow 0$.


\newpage

\section{Exciton Generation with Perpendicular Electric Field}
\label{appendix:exciton_generation}

In this Appendix we show that is possible to create interlayer electron-hole pairs in a heterojunction made of two semiconductor monolayers by using light with the electric field polarized normal to the layers.

The quantum geometric properties of interlayer excitons arise in part from the constituent monolayer band-structures. In order to minimize changes in these due to interlayer effects, we focus on heterostructures with weak tunnel coupling between the layers.
This could occur for two TMD semiconductors separated by one or more layers of hBN.  Moreover this is also the case for some specific materials, for example a heterojunction made of MoS$_2$ and WS$_2$ monolayers. The band structure of MoS$_2$ and WS$_2$ are well described by Hamiltonians of the form of Eq. \ref{MoS2 Hamiltonian 0} with appropriate parameters\cite{Xiao_2012}, yielding energies
\begin{equation}
\epsilon _n^{\pm}(\tau, \sigma ,{\bf k})= \frac {2 \tau \lambda _n} 2 \pm \sqrt { \left ( \delta _n - \frac {\sigma \tau \lambda_n} 2\right ) ^2 + v_n^2 k^2.
} \, \, \,
\end{equation}
The band structure for this heterostructure is essentially the superposition of the constituent semiconductor bands, shifted with respect to one another by an offset of $\Delta$=220 meV \cite{Kosmider_2013}. The optically minimal energy gap involves a hole in the valence band of WS$_2$ and an electron in MoS$_2$, with energy $\sim$ 1.45 eV.
In this structure, although the tunneling between the layers leaves the MoS$_2$ and WS$_2$ bands nearly unaltered, it does not vanish, allowing for the creation of interlayer electron-hole pairs\cite{Yu_2015, Wang_2017}.  These pairs can be created by sending light with the electric field polarized in the plane of the heterostructure; however this geometry generates several {\it intra-}layer excitons which could mask the physics of the interlayer excitons.  We therefore propose the use of light with the electric field perpendicular to the layers.

Suppose the two semiconductors, $n=\pm 1$, are separated by a distance $d$. The incident electromagnetic field has the electric field ${\mathbfcal E}$ polarized perpendicular to the bilayer and the momentum parallel to the layers. Since the wavelength of the light is very long compared to the lattice spacing in each layer, we treat the electric field as uniform in space and oscillatory in time. This gives a time dependent perturbation of the form

\begin{equation}
V(t)=e {\mathbfcal E} \frac d 2  \sum _{\tau , \sigma} { \left ( \hat N _{1,c} (\tau,\sigma)- \hat N _{1,v} (\tau,\sigma) -\hat N _{-1,c} (\tau,\sigma)+ \hat N _{-1,v} (\tau,\sigma) \right ) }\cos {\omega t} \equiv {\hat \Phi}   \cos {\omega t}
\end{equation}
with
\begin{equation}
\hat N _{n,c} (\tau,\sigma)= \sum _{ {\bf k} } c_n ^\dag (\tau,\sigma,{\bf k}) c_n  (\tau,\sigma,{\bf k})  \, \, \, {\rm and}  \, \,
\hat N _{n,v} (\tau,\sigma)= \sum _{ {\bf k} } v_n ^\dag (\tau,\sigma,{\bf k}) v_n  (\tau,\sigma,{\bf k}).
\end{equation}
Here $c_n ^\dag (\tau,\sigma,{\bf k})$ ($v_n ^\dag (\tau,\sigma,{\bf k})$)  creates an electron (hole) in layer $n$, in the valley $\tau$, with spin $\sigma$ and  momentum ${\bf k}$.  We are interested in the optical absorption that conserves both spin and valley index and therefore in what follows we omit these indices unless necessary.

The absorption rate can be written formally using Fermi's Golden Rule,
\begin{equation}
\Gamma = 2 \pi \sum _{f} | \langle f| {\hat \Phi} |i \rangle |^2 \delta (E_f-E_i-\hbar \omega) \, \, ,
\end{equation}
with $|j \rangle$ and $E_j$ eigenstates and eigenvalues respectively of the system.
We assume the system is initially in the ground state $|i \rangle$.  If we completely ignore interlayer tunneling then $\Gamma$ vanishes since
${\hat \Phi}$ does not change particle number in each layer. For this reason we include tunneling perturbatively. To first order,
\begin{equation}
|i \rangle   \cong |i \rangle_0 + \sum_{j \ne i} \frac { _0 \!   \langle j| {\hat H}_T |i \rangle _0}{E_j^{(0)}-E_i^{(0)}} |j \rangle_0,
\end{equation}
where  $ |i \rangle _0$ and $E_i^{(0)}$ represent eigenvectors and eigenvalues in the absence of tunneling.
The creation operators in the presence of tunneling become
\begin{eqnarray}
v_n ^\dag  ({\bf k}) & \rightarrow &v_{n}^\dag  ({\bf k}) + \frac {t_{vv}  }
{E_{-n}^- ({\bf k}=0)-E_{n}^- ({\bf k}=0) }
v^+_{-n}  ({\bf k})+
\frac {t_{cv}}
{E_{-n}^+ ({\bf k}=0)-E_{n}^- ({\bf k}=0) }
c_{-n}  ({\bf k}),\nonumber \\
c_n ^\dag   ({\bf k})& \rightarrow & c_{n}^\dag  ({\bf k})+ \frac {t_{cc}}
{E_{-n}^+ ({\bf k}=0)-E_{n}^+ ({\bf k}=0) }
c^+_{-n}  ({\bf k})  +
\frac {t_{cv}}
{E_{-n}^- ({\bf k}=0)-E_{n}^+ ({\bf k}=0) }
v_{-n}    ({\bf k}),
\label{pert}
\end{eqnarray}
where we have neglected  the wavevector dependence in the denominators.  In these expressions
${t_{cv}}$, $t_{cc} $ and $ t_{vv}$ are real parameters that can be obtained from {\it ab initio} calculations. For
MoS$_2$-WS$_2$ heterojunctions they are of order several  meV \cite{Yu_2015,Wang_2017}.

The non-vanishing contributions from the
number operators acting on the initial state of the system (vacuum) are
\begin{eqnarray}
\hat N _{n,c}  |i \rangle & = &
\sum _{{\bf k}} c^\dag _{n} ({\bf k}) \frac {t_{cv}}
{E_{-n}^+ ({\bf k}=0)-E_{n}^- ({\bf k}=0) }
v_{-n} ^\dag ({\bf k})  |i \rangle, \nonumber \\
\hat N _{n,v}  |i \rangle& = &
\sum _{{\bf k}} v^\dag _{n} ({\bf k}) \frac {t_{cv}}
{E_{-n}^- ({\bf k}=0)-E_{n}^+ ({\bf k}=0) }
c_{-n} ^\dag ({\bf k}) |i \rangle.
\end{eqnarray}

Taking into account the spin and valley indices, for the MoS$_2$-WS$_2$ system the resulting optical absorption becomes
\begin{eqnarray}
{\hat \Gamma} \approx 2 	\pi ( e {\mathbfcal E}  d)^2 t_{cv}^2 \sum _{\bf k}
&& \biggr[ \frac 1 {(\delta _1+\delta_{-1}-\lambda _{-1}-\Delta + O(k^2) )^2} \delta (\delta _1+\delta_{-1}-\lambda _{-1}-\Delta + O(k^2) -\hbar \omega)  \nonumber \\
&& +\frac 1 {(\delta _1+\delta_{-1}+\lambda _{-1}-\Delta + O(k^2))^2} \delta (\delta _1+\delta_{-1}+\lambda _{-1}-\Delta + O(k^2) -\hbar \omega)  \nonumber \\
&& +\frac 1 {(\delta _1+\delta_{-1}-\lambda _{1}+\Delta + O(k^2) )^2} \delta ( \delta _1+\delta_{-1}-\lambda _{1}+\Delta + O(k^2) -\hbar \omega)  \nonumber \\
&& +\frac 1 {( \delta _1+\delta_{-1}+\lambda _{1}+\Delta + O(k^2) )^2} \delta (  \delta _1+\delta_{-1}+\lambda _{1}+\Delta + O(k^2) -\hbar \omega) \biggl]  \, \, \, .
\end{eqnarray}

In this analysis we have assumed that the two materials are perfectly stacked upon one another. This is possible because  MoS$_2$ and WS$_2$  have nearly the same lattice parameter (mismatch less that 0.0013$\%$). Interestingly, if one of the semiconductors is rotated with respect to the other, the twist induces a modulation of the hopping amplitude between the  semiconductor  layers, and the tunneling between  the layers involves a momentum transfer\cite{Brey_2014b,Wang_2017}.  Thus a measurement of the exciton absorption resonance as a function of twist angle allows one to directly observe the exciton dispersion \cite{Yu_2015}. It is interesting to note that, for small rotation angles, only three transfer momentum are relevant for each valley so that the final exciton state generated in the process we describe will in fact be an equal, coherent admixture of states with momenta related by C$_6$ symmetry. We assume, as in other such heterostructures, that this state quickly relaxes to, up to thermal fluctuations, the ground state of the exciton.





\end{document}